\documentclass[preprint]{aastex62}

\usepackage{acronym}
\usepackage{booktabs}
\usepackage{dcolumn}
\usepackage{graphicx}
\usepackage{amsmath,amsfonts,amssymb}
\usepackage{mathrsfs}
\usepackage[normalem]{ulem}
\graphicspath{{./}{figures/}}

\shorttitle{Axion Minihalos \& Highly Magnified Stars}
\shortauthors{Dai \& Miralda-Escud\'{e}}

\newcommand{\refeq}[1]{Eq.~(\ref{eq:#1})}          
          
\newcommand{\reffig}[1]{Figure~\ref{fig:#1}}          
\newcommand{\refsec}[1]{Section~\ref{sec:#1}}
\newcommand{\refapp}[1]{Appendix~\ref{app:#1}}

\newcommand{\be}{\begin{equation}}
\newcommand{\ee}{\end{equation}}
\newcommand{\ba}{\begin{eqnarray}}
\newcommand{\ea}{\end{eqnarray}}

\newcommand{\bfx}{\boldsymbol{x}}
\newcommand{\bfy}{\boldsymbol{y}}
\newcommand{\bfq}{\boldsymbol{q}}
\newcommand{\bfalp}{\boldsymbol{\alpha}}
\newcommand{\bfell}{\boldsymbol{\ell}}

\newcommand{\bfr}{\boldsymbol{r}}
\newcommand{\deltwod}{\delta_{\rm 2d}}

\newcommand{\barSigma}{\overline{\Sigma}}

\def\VEV#1{\left\langle #1 \right\rangle}
\def\rmd{\mathrm{d}}

\begin{document}

\title{Gravitational Lensing Signatures of Axion Dark Matter Minihalos in Highly Magnified Stars}

\correspondingauthor{Liang Dai}
\email{ldai@ias.edu}

\author[0000-0003-2091-8946]{Liang Dai}
\affiliation{School of Natural Sciences, Institute for Advanced Study,
1 Einstein Drive,
Princeton NJ 08540, USA}

\author[0000-0002-2316-8370]{Jordi Miralda-Escud\'e}
\affiliation{School of Natural Sciences, Institute for Advanced Study,
1 Einstein Drive,
Princeton NJ 08540, USA}
\affiliation{Institut de Ci\`encies del Cosmos, Universitat de Barcelona, Mart\'\i\ i Franqu\`es 1, 08028 Barcelona, Catalonia}
\affiliation{Instituci\'o Catalana de Recerca i Estudis Avan\c cats, Passeig Llu\'\i s Companys 23, 08010-Barcelona, Catalonia.}



\begin{abstract}

Axions are a viable candidate for Cold Dark Matter (CDM) which should generically form minihalos of sub-planetary masses from white-noise isocurvature density fluctuations if the Peccei-Quinn phase transition occurs after inflation. Despite being denser than the larger halos formed out of adiabatic fluctuations from inflation, axion minihalos have surface densities much smaller than the critical value required for gravitational lensing to produce multiple images or high magnification, and hence are practically undetectable as lenses in isolation. However, their lensing effect can be enhanced when superposed near critical curves of other lenses. We propose a method to detect them through photometric monitoring of recently discovered caustic transiting stars behind cluster lenses, under extreme magnification factors $\mu \gtrsim 10^3$--$10^4$ as the lensed stars cross microlensing caustics induced by intracluster stars. For masses of the first gravitationally collapsed minihalos in the range $\sim 10^{-15}$--$10^{-8}\,h^{-1}\,M_\odot$, we show that axion minihalos in galaxy clusters should collectively produce subtle surface density fluctuations of amplitude $\sim 10^{-4}$--$10^{-3}$ on projected length scales of $\sim 10$--$10^4\,$AU, which imprint irregularities in the microlensing light curves of caustic transiting stars. We estimate that, inside a cluster halo and over the age of the Universe, most of these minihalos are likely to avoid dynamic disruption by encounters with stars or other minihalos. 

\end{abstract}


\bigskip
$~$

\section{Introduction}
\label{sec:intro}

Gravitational lensing can reveal self-gravitating objects containing only dark matter (DM), which are difficult to detect by other means because they have expelled or have never accreted baryons. Most DM theories predict the presence of these objects on mass scales much smaller than that of galaxies. Direct detection of these structures would provide crucial clues to the physical origin of DM. A common problem is that these structures are predicted to have surface densities well below the critical value, so they by themselves cannot act as gravitational lenses capable of creating multiple images or inducing substantial image magnifications and distortions, and their effects are often unnoticeable.

However, when these small-scale objects are superposed on a large-scale lens near lensing critical curves, very small surface density perturbations may result in large observable changes~\citep{minor2017robust, Dai:2018mxx}. Recently, the first highly magnified stars~\citep{1991ApJ...379...94M} have been discovered in the field of massive clusters of galaxies~\citep{Kelly:2017fps, Chen:2019ncy, Kaurov:2019alr}. Near the critical curve of the cluster, individual superluminous stars in the background become visible to our most powerful telescopes when magnified by a factor $\sim 10^2$--$10^3$. Microlensing by intracluster stars in the lensing cluster introduces fast variability in the magnification~\citep{Venumadhav:2017pps, Diego:2017drh, Oguri:2017ock}, turning highly magnified stars into a probe not only of the abundance and mass function of intracluster stellar objects, but also of any small-scale density inhomogeneity in the DM.

Axions are a promising DM candidate produced by non-thermal mechanisms in the early Universe. For a famous example, the QCD axion is motivated by the Peccei-Quinn mechanism~\citep{peccei1977constraints}, which explains the absence of a CP violating term in the Quantum Chromodynamics (QCD) lagrangian~\citep{weinberg1978new, wilczek1978problem, kim1979weak, shifman1980can, dine1981simple} as implied by the null value of the neutron electric dipole moment~\citep{Afach:2015sja}. Generic axions can arise from other spontaneous broken U(1) symmetries~\citep{chikashige1981there, gelmini1981left, wilczek1982axions, berezhiani1990theory, jaeckel2014family} or from low-energy effective theories emerging from string theory~\citep{witten1984some, conlon2006qcd, cicoli2012type, georgi1981grand, choi2009accions, dias2014quest}, and may also account for the DM. 

Cosmological models involving axions as the DM predict the existence of very dense, low-mass dark matter halos through gravitational instability~\citep{hogan1988axion,kolb1994large} if the Peccei-Quinn symmetry breaking occurs after inflation. However, the surface density of these minihalos would be too low to make them generally detectable through gravitational lensing. Despite a number of proposed lensing methods to detect small-scale axion structures~\citep{kolb1996femtolensing, Fairbairn:2017dmf, fairbairn2018structure}, so far the range of axion minihalo densities and masses do not seem accessible with any foreseeable observational techniques.

In this paper we propose the first realistic astrophysical method to detect axion minihalos through gravitational lensing. Highly magnified stars lensed by clusters of galaxies can have apparent fluxes affected by the presence of axion minihalos when they are crossing one of the micro-caustics produced by intracluster stars. While the intracluster stars disrupt the cluster critical curve into a network of micro-critical curves on angular scales of $\sim 10\, \mu{\rm as}$, axion minihalos having sub-planetary masses and solar system sizes would produce finer magnification variations on much smaller scales of $\sim 10^2$--$10^3$ nano-arcseconds. When crossing a micro-caustic, the observed light curve of a highly magnified star should be altered by surface density fluctuations caused by axion minihalos inside the cluster lens. This paper aims to predict the lensing impacts of axion minihalos on highly magnified stars, which can be monitored through dedicated observations with our most powerful space-borne or ground-based optical and infrared telescopes~\citep{windhorst2018observability, diego2019universe}.

We outline this work as the following. In \refsec{iso}, we start with a review on primordial isocurvature density fluctuations expected to arise from a Peccei-Quinn phase transition in axion DM scenarios. Then in \refsec{sizedensity}, we derive the physical properties of axion minihalos that collapse from primordial isocurvature density fluctuations, and show that they are significantly denser and more compact than halos of comparable masses expected in the standard cosmology. In \refsec{massfun}, we discuss the mass function of axion minihalos. In \refsec{powerspec2d}, we calculate the size of surface mass density fluctuations due to the cumulative effect of many substructure minihalos along the line of sight through the cluster halo. In \refsec{disruption}, we estimate possible dynamic disruption of axion minihalos by mutual encounters or encounters with stars. In \refsec{microlensing}, we study the lensing signatures of axion minihalos imprinted on highly magnified stars, and show that sizable irregularities are expected in their light curves during micro-caustic crossings. Additional discussion is presented in \refsec{discuss}, and we conclude in \refsec{concl}. We discuss some technical details in a few Appendices. We use the conventional notation in cosmology that many physical quantities are expressed in a way that scales with the dimensionless Hubble parameter $h$.

\section{Isocurvature density fluctuations in axion cosmology}
\label{sec:iso}

Axions are a viable particle candidate for non-thermal DM, namely DM that never interacted substantially with the baryonic matter in the Universe since it was created. The axion DM hypothesis is that a global U(1) symmetry of a complex scalar field $\varphi$ was spontaneously broken in the very early Universe~\citep{peccei1977constraints} and gave rise to the axion particle $\phi$ as the Goldstone mode associated with the angle of $\varphi$. If the global U(1) symmetry breaking occurs after inflation during the era of radiation domination, causally disconnected Hubble patches settle down to different vacuum expectation values of $\phi$.

The axion can acquire a non-zero mass $m_a$ through non-perturbative effects such as instantons~\citep{gross1981qcd} which can tilt the potential so that it is no longer flat with respect to $\phi$. The axion mass is in general temperature dependent, which induces fluctuations of order unity in the axion field energy density as the field configuration oscillates around the potential minimum $\phi=0$. This, however, does not occur immediately after the axion acquires its mass because the Hubble timescale is shorter than $\hbar/(m_a\,c^2)$ in the beginning and the axion is dynamically ``frozen''. During this time, the field configuration is smoothed over the instantaneous horizon scale by the Kibble mechanism~\citep{kibble1976topology}.

The situation changes when the cosmic expansion eventually slows down to be comparable to the axion field oscillation time scale, $3\,H(t_0) \simeq m_a(t_0)\,c^2/\hbar$. The axion field, which is coherent on the horizon scale at this epoch, $\lambda_H(t_0) \sim c/H(t_0)$, starts to oscillate around $\phi=0$ and to contribute to the cosmic matter density budget. This generates isocurvature perturbations in addition to the inflationary adiabatic perturbations. The QCD axion starts to oscillate when the Universe cools to a temperature $T_0 = T(t_0) \simeq 1.3\,{\rm GeV}$~\citep{Buschmann:2019icd}, and the axion mass evolves with the temperature from $m_a(T_0) \sim 6 \times 10^{-9}\,{\rm eV}$ to $m_a = 2.5\times 10^{-5}\,{\rm eV}$ at zero temperature~\citep{di2016qcd, Buschmann:2019icd}, although the exact numbers are subject to theoretical uncertainties~\citep{borsanyi2016calculation}. A generic axion particle can have a different mass for which the onset of field oscillation is at a different epoch.


Assuming Gaussian statistics, the total power spectrum for the matter overdensity is the sum of the standard inflationary adiabatic power spectrum $P_{\rm cdm}$ and the axion isocurvature power spectrum $P_{\rm iso}$~\footnote{Throughout the paper we adopt the simplifying terminology ``isocurvature fluctuation'' to mean the perturbation in the ratio of the axion density to total density, with a constant baryon-to-photon ratio, which should be distinguished from other forms of isocurvature modes.}. At the initial time $t_0$, the latter can be modeled as a white-noise power spectrum~\citep{hogan1988axion, fairbairn2018structure} $P_{\rm iso}(k)  = \Theta(k_0 - k)\,24\,\pi^2/(5\,k^3_0)$, where $k_0=a(t_0)\,H(t_0)/c$ is a cutoff comoving wave number associated with the horizon scale at $t_0$. Requiring a DM overdensity variance of order unity on the scale $k_0$ at $t_0$ fixes the normalization of $P_{\rm iso}$. The total amount of {\it axion} mass enclosed within a spherical volume of a comoving radius $\pi/k_0$ defines a characteristic mass scale
\ba
\label{eq:M0}
M_0 = (4\,\pi/3)\,\left( \pi/k_0 \right)^3\,\bar{\rho}_{a0},
\ea
where $\bar{\rho}_{a0}$ is the present day mean density in axions. This sets the mass scale of the first collapsed objects from the isocurvature perturbations. For the QCD axion, we obtain $M_0 \simeq 5 \times 10^{-10}\,M_\odot$ based on the axion mass parameterization used in \cite{Buschmann:2019icd} and the the effective number of relativistic degrees of freedom presented by \cite{husdal2016effective} (see \refapp{M0}). After accounting for a factor of $4\,\pi^4/3\approx 130$ larger in our definition, this number is compatible with the characteristic masses for the first gravitationally collapsed axion minihalos quoted in other references~\citep{Davidson:2016uok, hardy2017miniclusters, fairbairn2018structure}. We therefore estimate that the typical mass for the first collapsed axion minihalos is $\sim 0.01\,M_0$, and is $\sim 5 \times 10^{-12}\,M_\odot$ for QCD axions. For generic axion DM, a wide range of values for $M_0$ are possible.


The white noise model is not likely to be a good approximation around the cutoff scale $k_0$, where the axion field fluctuations are subject to complicated nonlinear dynamics and can be significantly non-Gaussian~\citep{kolb1996femtolensing}. Numerical simulations find intricate axion clumps on scales shorter than $k_0^{-1}$~\citep{Vaquero:2018tib, Buschmann:2019icd}. Unrelated to gravitational instability, these clumps emerge soon after the onset of field oscillation at $t_0$ as a result of complicated Klein-Gordon dynamics, with mass scales as much as four orders of magnitude smaller than the analytic estimate $\sim 0.01\,M_0$ for the smallest gravitationally collapsed minihalos~\citep{Vaquero:2018tib, Buschmann:2019icd}. These clumps are likely to be contained within the first gravitationally collapsed minihalos, and we distinguish them from axion minihalos, which are the focus of study throughout this paper. The Gaussian white-noise model for minihalos should work well in the shot-noise regime $k\ll k_0$, regardless of the formation of smaller clumps.

With favorable parameter values for the axion cosmology, $k_0$ typically translates into a length scale many orders of magnitude smaller than the range of scales probed through the Cosmic Microwave Background anisotropies and the large-scale clustering of galaxies. Therefore, the isocurvature modes that dominate over the adiabatic modes enter the horizon well before the epoch of radiation-matter equality at $z_{\rm eq}\simeq 3400$, so they have wave numbers $k \gg k_{\rm eq} = a(z_{\rm eq})\,H(z_{\rm eq})/c$. Analytic and numerical calculations of linear perturbations have shown that sub-horizon isocurvature matter modes start to grow substantially only when $z > z_{\rm eq}$~\citep{efstathiou1986isocurvature}. Adiabatic modes behave differently during the era of radiation domination: upon entering the horizon, the oscillation of the photon-baryon fluid as an acoustic wave generates an initial peculiar velocity in the DM, leading to a comoving displacement of the DM that grows logarithmically with the scale factor and is responsible for the shape of the linear CDM power spectrum at $k> k_{\rm eq}$. In comparison, isocurvature modes induce no such initial peculiar velocities, and grow only through the self gravitation of the DM after horizon entry and the photon-baryon acoustic oscillation (see \refapp{matter} for more details).

Since the white noise isocurvature power spectrum is only meant to be a crude model, we adopt the approximation that all the isocurvature modes grow by the same factor, equal to the usual $\Lambda$CDM growth factor $D_+(z)$. We neglect the growth of isocurvature modes at $z < z_{\rm eq}$, noting that this small growth is in any case a nearly constant factor over the relevant range of scales that can be reabsorbed into the definition of $k_0$. We can then write the linear power spectrum for the total matter overdensity (including both isocurvature and adiabatic modes, but neglecting the gravitational influence of baryons on the isocurvature component) at redshift $z$ as
\ba
\label{eq:Pm_tot}
P_m(k, z) & = & P_{\rm cdm}(k)\,D^2_+(z) + \Theta\left(k_0 - k\right)\,\frac{24\,\pi^2}{5\,k^3_0}\,\left( \frac{D_+(z)}{D_+(z_i)} \right)^2\,\left(\frac{1+z_{\rm eq}}{1+z_i} \right)^2.
\ea
This is valid for $z < z_{\rm eq}$, where $z_i$ is an arbitrarily chosen reference redshift during the era of matter domination satisfying $1 \ll z_i \ll z_{\rm eq}$. 

We note that any primordial {\it adiabatic} fluctuations of large amplitude and on small scales would be subject to stringent constraints on entropy production from acoustic damping in the primordial plasma~\citep{Jeong:2014gna, Inomata:2016uip}. Since modes of $k \sim 10^2$--$10^8\,h\,{\rm Mpc}^{-1}$ enter the horizon well before the epoch of radiation-matter equality, they would have rapidly dissipated. By contrast, large CDM isocurvature fluctuations are allowed because density fluctuations during the radiation era $z \gg z_{\rm eq}$ are highly suppressed. Neither are these axion isocurvature modes subject to Big Bang Nucleosynthesis constraints, which apply to {\it baryonic} isocurvature modes of $k \lesssim 10^9\,h\,{\rm Mpc}^{-1}$~\citep{Inomata:2018htm}.

\begin{figure}
  \includegraphics[width=0.98\columnwidth]{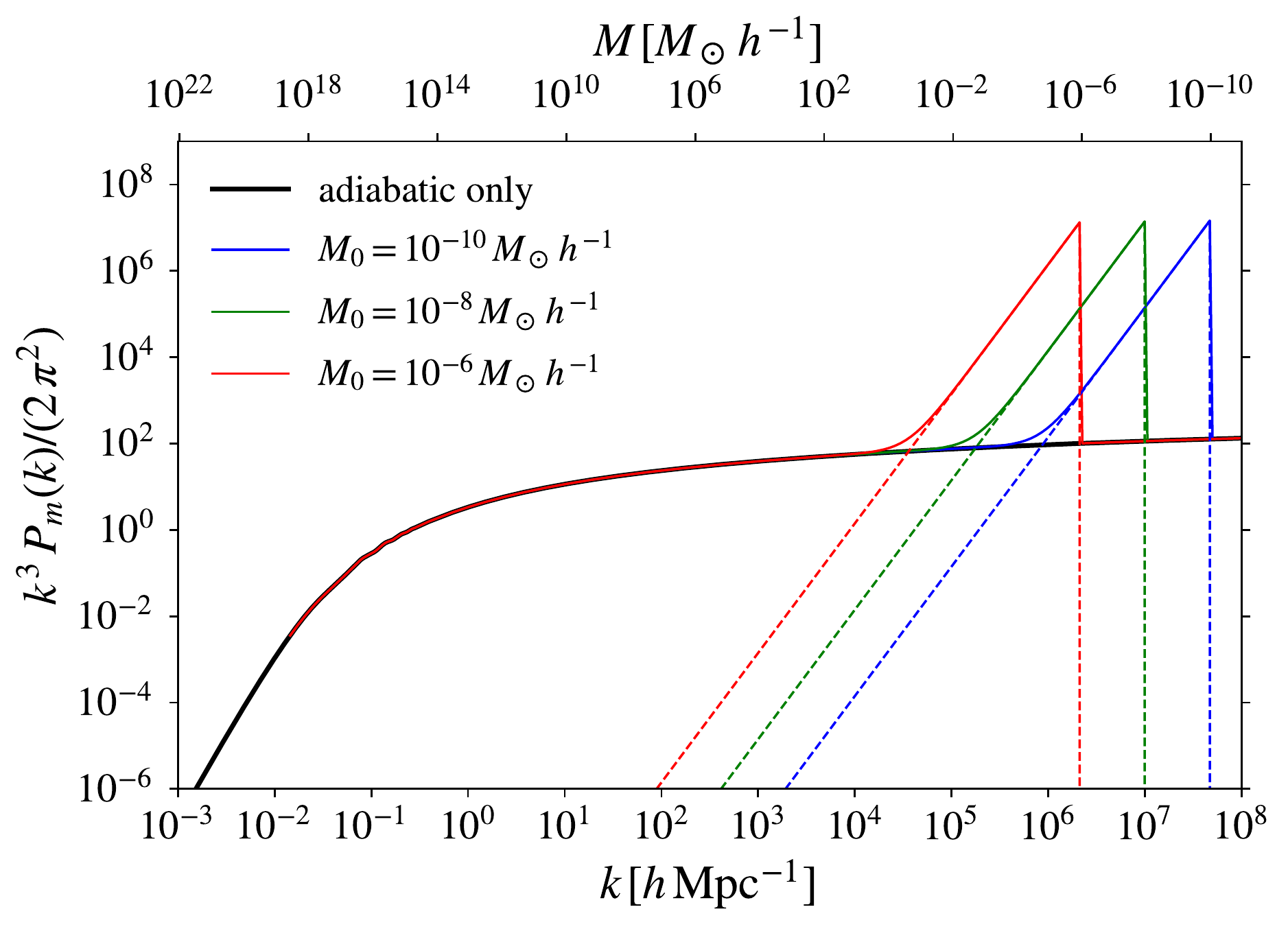}
  \caption{Linear power spectrum of matter overdensity $P_m(k)$ at $z=0$ assuming all DM is made of axions. The black solid curve shows the standard adiabatic power spectrum, and the colored dashed curves show isocurvature contributions corresponding to various values of $M_0$. Only isocurvature modes of $k > k_{\rm eq}$ which enter the horizon prior to radiation-matter equality are shown. For reference, $k$ is converted into a mass scale $M=\bar{\rho}_{a0}\,(4\pi/3)\,(\pi/k)^3$ along the top axis. }
  \label{fig:Pm}
\end{figure}

In \reffig{Pm}, we plot the total matter overdensity power spectrum linearly extrapolated to $z=0$ for $M_0$ in the range $10^{-10}$--$10^{-6}\,M_\odot/h$. On comoving scales shorter than the inverse of $k \sim 10^4$--$10^8\,h/{\rm Mpc}$, the isocurvature contribution can dominate the total power. This means that isocurvature perturbations lead to earlier gravitational collapse on those small scales. In fact, on scales $k\sim k_0$, fluctuations are initially of order unity and they reach non-linear collapse near the epoch of radiation-matter equality. Note that the sharp reduction of power at scale $k_0$ in \reffig{Pm} is expected to be smoothed once axion dynamics on scales shorter than $k_0$ are accurately calculated.

On sufficiently small scales, gravitational instability is prohibited by the quantum degeneracy pressure of the axion field. The comoving wave number of the effective Jeans scale is given by~\citep{fairbairn2018structure}
\ba
\label{eq:kJ}
k_J  = (16\,\pi\,G\,a\,\bar{\rho}_{a0})^{1/4} \,\left(m_a/\hbar\right)^{1/2} = 6.7\times 10^{10}\,{\rm Mpc}^{-1}\,(1+z)^{-1/4}\,\left(\frac{\Omega_m\,h^2}{0.12}\right)^{1/4}\,\left( \frac{m_a}{10^{-4}\,{\rm eV}} \right)^{1/2}.
\ea
Structure formation becomes efficient at $z < z_{\rm eq} \simeq 3400$, and the expected QCD axion has a mass in the range $m_a \sim 10^{-6}$--$10^{-4}\,$eV at zero temperature, so \refeq{kJ} suggests that quantum degeneracy pressure is negligible for the gravitational instability of the isocurvature modes shown in \reffig{Pm}.

\section{Size and density of axion minihalos}
\label{sec:sizedensity}

During the epoch of matter domination, collapsed objects form after the matter overdensity amplitude reaches a value near unity. Around the epoch of radiation-matter equality, the first collapsed objects form on a comoving scale $\sim 1/k_0$ with masses $\sim 0.01\,M_0$. Similar to that in the standard cosmology, hierarchical assembly of collapsed objects proceeds in a bottom-up fashion toward larger length scales (and smaller comoving wave numbers). Increasingly larger halos are built up from mergers of smaller ones and accretion of smooth matter, and they develop an internal structure of orbiting satellite halos, with at least a fraction of them surviving dynamical disruption. This structure may have multiple levels of orbiting satellites within orbiting satellites. The hierarchical process takes place first during $z_{\rm eq} > z \gtrsim 20$ on very small scales, where the white-noise isocurvature power spectrum dominates. Then, at $z \lesssim 20$, the inflationary adiabatic power spectrum is dominant, and structure formation proceeds in a way identical to that of the standard CDM cosmology, except that a fraction of the DM is clumped in the form of vast numbers of very small and dense orbiting minihalos.

Consider the collapse of a spherical region at redshift $z_{\rm coll}$, which has a comoving Lagrangian radius $R$ and an enclosed mass $M = (4\pi/3)\,\bar{\rho}_{a0}\,R^3$. Exact spherical symmetry in collapse is unrealistic for predicting the correct density profile of the collapsed halos. Numerical studies of collapse from Gaussian random overdensities using N-body simulations suggest that halos formed through hierarichical assembly in general follow the phenomenological Navarro-Frenk-White (NFW) density profile~\citep{1996ApJ...462..563N, navarro1997universal} $\rho(r) = \rho_s/[(r/r_s)(1+r/r_s)^2]$, which is defined by a scale density $\rho_s$ and a scale radius $r_s$.

The literature on CDM halos has commonly used three quantities to characterize an NFW halo: the virial mass $M_{200}$, the virial radius $r_{200}$, and the concentration parameter $c_{200} =r_{200}/r_s$. Instead, we treat $\rho_s$ and $r_s$ as fundamental parameters which are closely tied to the time of halo collapse. Numerical studies suggest a ``two step'' picture for the formation and evolution of halos (especially at low peak heights): upon collapse, the halo first undergoes a phase of rapid growth to reach a typical concentration factor $c_\star \simeq 4$~\citep{zhao2003mass, zhao2009accurate}; after that, halo growth slows down while its inner profile stabilizes~\citep{bullock2001profiles, ludlow2013mass}. During the second phase, which is often referred to as {\it pseudo evolution}, $\rho_s$ and $r_s$ hardly change if measured in proper units, while the virial mass and the virial radius still increase due to mass accretion onto the outer parts. The picture of pseudo evolution applies particularly well to isolated halos that grow slowly from the low-density surrounding material, which are typically of low mass compared to the characteristic collapsing mass at each epoch. Using $\rho_s$ and $r_s$ is convenient for quantifying the gravitational lensing effect; they characterize the halo's inner region, which is much denser and more compact than the entire virialized region at high halo concentration. We therefore describe NFW halos using $\rho_s$ and $r_s$, assumed to be stationary from the onset of the pseudo-evolution phase.

The linear matter overdensity smoothed over a scale $R$ has a variance $\sigma^2(M, z) = \int\,\rmd\ln k\,$ $[k^3\,P_m(k, z)]/(2\pi^2)\, |W(k, R)|^2$, where $W(k, R)=3\,(kR)^{-3}\,[\sin(kR) - (kR)\,\cos(kR)]$ is the spherical top-hat window function. At each redshift, one can derive the mass scale $M$ and the comoving radius scale $R=R(M)$ that collapse from any given peak height $\nu(M, z) = \delta_c/\sigma(M, z)$, where $\delta_c = 1.686$ is the threshold overdensity during the era of matter domination. Halos of mass $M$ collapse over a broad range of redshifts, but we can define a universal {\it median} collapse redshift corresponding to a fluctuation with $\nu=\nu_{\rm med}=0.67$. For any $M$, a characteristic mass-dependent collapse redshift $z_{\rm coll}(M)$ follows by solving $\nu_{\rm med}=\delta_c/\sigma(M, z_{\rm coll})$. When isocurvature fluctuations dominate the overdensity power spectrum in \refeq{Pm_tot}, the median collapse redshift is derived to be
\ba
1 + z_{\rm coll}(M) = \left( 18/5\,\pi^2 \right)^{1/2}\,\left( \nu_{\rm med}/\delta_c\right)\,\left( 1 + z_{\rm eq} \right)\,\left( M_0/M \right)^{1/2} = 0.24\,\left( 1 + z_{\rm eq} \right)\,\left( M_0/M \right)^{1/2}.
\ea
To estimate the NFW fundamental parameters, we assume that the collapsed NFW halo has a universal concentration parameter $c_\star=r_{200}/r_s=4$, its enclosed mass within $r_{200}$ equals $M$, and the mean density within $r_{200}$ is 200 times the cosmic mean $\bar{\rho}_a(z_{\rm coll})=\bar{\rho}_{a0}\,(1+z_{\rm coll})^3$. For $c_\star=4$, we find $\rho_s \approx 5271\,\bar{\rho}_a(z_{\rm coll})$, $r_s \approx 0.0265\,[M/\bar{\rho}_a(z_{\rm coll})]^{1/3}$, and $M \approx 10.2\,\rho_s\,r^3_s$. When the white-noise isocurvature power spectrum dominates, the resultant halos, which we refer to as minihalos, obey the following scaling relations with the mass scale $M$,
\be
\rho_s(M) \approx 0.24\,M_\odot\,h^2/{\rm pc}^3\,\left( \frac{\nu}{\nu_{\rm med}} \right)^3\,\frac{\Omega_a}{0.3}\,\left( \frac{1+z_{\rm eq}}{3400} \right)^3\,\left(\frac{M_0}{10^{-10}\,M_\odot/h}\right)^{3/2}\,\left(\frac{M}{10^{-6}\,M_\odot/h}\right)^{-3/2},
\ee
and
\be
\label{eq:rs}
r_s(M) \approx 1530\,h^{-1}{\rm AU}\,\left( \frac{1+z_{\rm eq}}{3400}\, \frac{\nu}{\nu_{\rm med}} \right)^{-1}\,\left(\frac{\Omega_a}{0.3}\right)^{-1/3}\,\left(\frac{M_0}{10^{-10}\,M_\odot/h}\right)^{-1/2}\,\left(\frac{M}{10^{-6}\,M_\odot/h}\right)^{5/6}.
\ee

It is useful at this point to compare the characteristic {\it surface} density of axion minihalos, $\rho_s(M)\, r_s(M)$, with the critical surface density in gravitational lensing, required for an isolated lens to produce multiple images and high magnification. The critical surface density is given by $\Sigma_{\rm crit}=c^2/(4\pi\,G\,D_{\rm eff})$, where $D_{\rm eff}=D_L\,D_{LS}/D_S$ for angular diameter distances $D_L$ to the lens, $D_S$ to the source, and $D_{LS}$ from the lens to the source. The typical lensing convergence of an axion minihalo near the scale radius, $\kappa_s(M) \approx \rho_s(M)\,r_s(M)/\Sigma_{\rm crit}$, is tiny,
\be
\label{eq:kappaM}
\kappa_s(M) \approx 7\times 10^{-7}\,\frac{D_{\rm eff}}{1\,{\rm Gpc}}\, \left( \frac{\nu}{\nu_{\rm med}} \right)^2\,\left(\frac{\Omega_a}{0.3}\right)^{2/3}\,\left( \frac{1+z_{\rm eq}}{3400} \right)^2\,\left(\frac{M_0}{10^{-10}\,M_\odot/h}\right)\,\left(\frac{M}{10^{-6}\,M_\odot/h}\right)^{-2/3}.
\ee
Even for dense minihalos with $M = 10^{-10}\,M_\odot/h$, acting as individual lenses at cosmological distances, $\kappa_s(M)$ is no greater than $\sim 10^{-3}$. Axion minihalos can therefore not produce substantial lensing effects by themselves, but we will show in this paper that they can be detected when acting in conjunction with a galaxy cluster lens and intracluster stars near a lensing critical curve.

\begin{figure}
  \includegraphics[width=0.98\columnwidth]{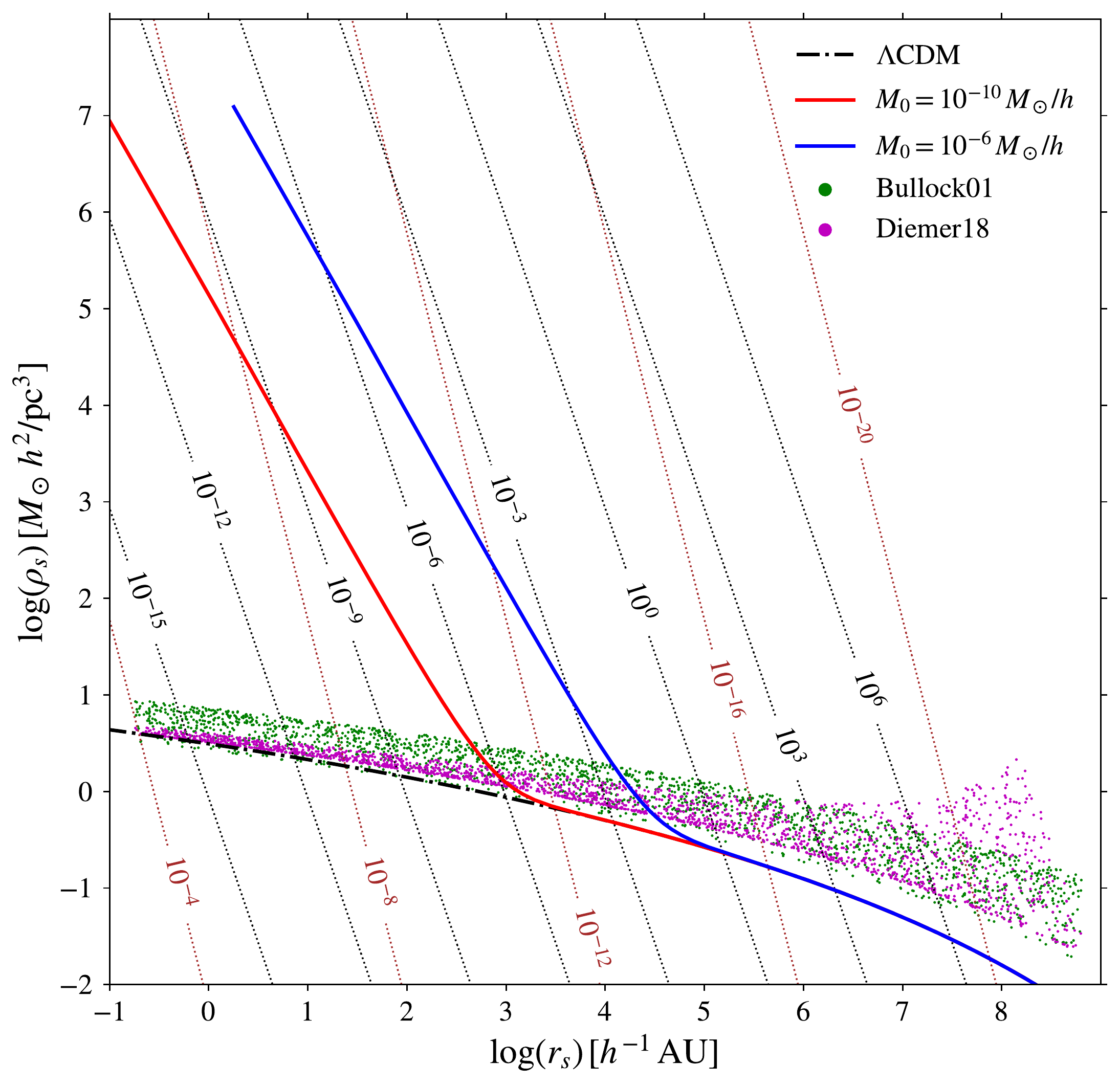}
  \caption{Scale radius $r_s$ versus scale density $\rho_s$ for NFW halos in standard or axion cosmologies. We scatter plot $r_s$ and $\rho_s$ for standard CDM halos drawn from $0 < z < 15$ and $-16 < \log(M_{200}\,h/M_\odot) < 10$ assuming the median halo concentration-mass relation from \cite{bullock2001profiles} (green dots) and from \cite{diemer2018accurate} (magenta dots). Curves show $r_s$ and $\rho_s$ calculated from the simple prescription that a spherical top-hat region enclosing mass $M$ with a median peak height $\nu = \nu_{\rm med} = 0.67$ collapses into an NFW halo with concentration $c_{200} = r_{200}/r_s = 4$ and mass $M$ within $r_{200}$. Black dash-dotted curve is for standard $\Lambda$CDM cosmology. Blue ($M_0=10^{-10}\,M_\odot/h$) and red ($M_0=10^{-6}\,M_\odot/h$) curves are computed using the axion cosmology power spectrum \refeq{Pm_tot}. Black dotted contours show $10.2\,\rho_s\,r^3_s$ in units of $M_\odot/h$. Brown dotted contours show the axion mass $m_a$ below which quantum degeneracy pressure is dynamically important, in units of ${\rm eV}\cdot h$.}
  \label{fig:rs_vs_rhos}
\end{figure}

In \reffig{rs_vs_rhos}, we check whether the above simple peak height prescription gives reasonably good estimates for the fundamental parameters $\rho_s$ and $r_s$ of halos at low masses. We first generate random NFW halos in the standard $\Lambda$CDM cosmology within a range of redshifts $0 < z < 15$ and a range of virial masses $-16 < \log(M_{200}\,h/M_\odot) < 10$ sampled from N-body simulations~\citep{diemer2018accurate}. Next, we calculate their $r_s$ and $\rho_s$ using a number of median concentration-mass relations from the literature, which are empirically calibrated to simulations. When we scatter plot the halos on the $(r_s,\,\rho_s)$ plane, they all lie within a squeezed band, which is particularly narrow at very small halo masses. This indicates that the notion of pseudo evolution is reasonably valid for low mass halos. For a comparison, we then analytically estimate $\rho_s$ and $r_s$ of median halos using the peak height prescription: we identify peaks with a median height $\nu = \nu_{\rm med}$ at various redshifts, and then calculate using the relations $\rho_s \approx 5271\,\bar{\rho}_a(z_{\rm coll})$ and $r_s \approx 0.0265\,[M/\bar{\rho}_a(z_{\rm coll})]^{1/3}$, where at any given collapse redshift $z_{\rm coll}$ we solve the top-hat mass $M$ from $\nu(M, z_{\rm coll}) = \delta_c/\sigma(M, z_{\rm coll})=\nu_{\rm med}$. On the $(r_s,\,\rho_s)$ plane, this produces a single curve parameterized by the collapse redshift $z_{\rm coll}$ or alternatively the top hat mass $M$. At small halo masses, this curve aligns reasonably well with the narrow band of halos identified from numerical simulations.

With this justification, hereafter we rely on the above peak height prescription to estimate halo size and density in axion cosmology. We show some sample curves in \reffig{rs_vs_rhos}. We see that in axion cosmology, sufficiently small halos can be many orders of magnitude denser than their $\Lambda$CDM counterparts~\citep{diemand2005earth, berezinsky2006destruction, diemand2006early}.

The inner region of the collapse axion halo has a typical de Broglie wavelength $\lambda_{\rm dB} =(2\,\pi\,\hbar)/(m_a\,\sigma)$, where we estimate velocity dispersion as $\sigma \sim (G\,\rho_s\,r_s)^{1/2}$. The quantum degeneracy pressure is dynamically unimportant if $\lambda_{\rm dB} \ll r_s$, which requires
\be
m_a >  \left[ \frac{(2\,\pi\,\hbar)^2}{G\,\rho_s\,r^4_s} \right]^{1/2} \approx 8\times 10^{-12}\,{\rm eV}\,h\, \left( \frac{\rho_s}{10^4\,M_\odot\,h^2\,{\rm pc}^{-3}} \right)^{-1/2}\,\left( \frac{r_s}{10^2\,h^{-1}\,{\rm AU}} \right)^{-2}.
\ee
Whenever this is valid, the axion halo can be modeled as a self-gravitating system made of classical collisionless particles (if axion self-interaction is negligible). In \reffig{rs_vs_rhos}, we overplot contours for this minimum axion mass, such that for any given value of $m_a$ halos located to the upper right side of the contour on the $(r_s,\,\rho_s)$ plane are {\it not} dynamically affected by the degeneracy pressure. For the mass range $m_a \sim 10^{-6}$--$10^{-4}\,$eV (at zero temperature) which brackets the theoretically favorable mass for the QCD axion, axion minihalos formed from isocurvature fluctuations are not affected by the degeneracy pressure.

\section{Mass function of axion minihalos}
\label{sec:massfun}

The Press-Schechter (PS) formalism~\citep{press1974formation} predicts that at a given time $t$ the collapsed halos have a mass function $n(M, t) = (\bar{\rho}_{m}(t)/M^3)\,\rmd f(M, t)/\rmd \ln M$. We define $n(M, t)$ to be the proper volume number density of halos per logarithmic interval of mass $M$. The mass fraction per logarithmic interval of mass is
\be
\label{eq:dfdlnM}
\frac{\rmd f(M,\,t)}{\rmd \ln M} = \sqrt{\frac{2}{\pi}}\,\nu(M, t)\,\exp\left(-\frac{\nu^2(M, t)}{2}\right)\,\left| \frac{\partial\ln\sigma(M, t)}{\partial\ln M} \right|.
\ee
Here $\bar{\rho}_m(t)$ is the mean matter density, and $\nu(M, t)=\delta_c/\sigma(M, t)$ is the peak height.

The PS mass function excludes small halos that have been incorporated into larger ones as subhalos. Some subhalos are disrupted, while others may survive for long. Both isolated halos and subhalos act as perturber lenses, so as far as gravitational lensing is concerned, \refeq{dfdlnM} sets a lower limit to the fraction of mass locked up in collapsed, dense objects.

\begin{figure}
  \includegraphics[width=0.98\columnwidth]{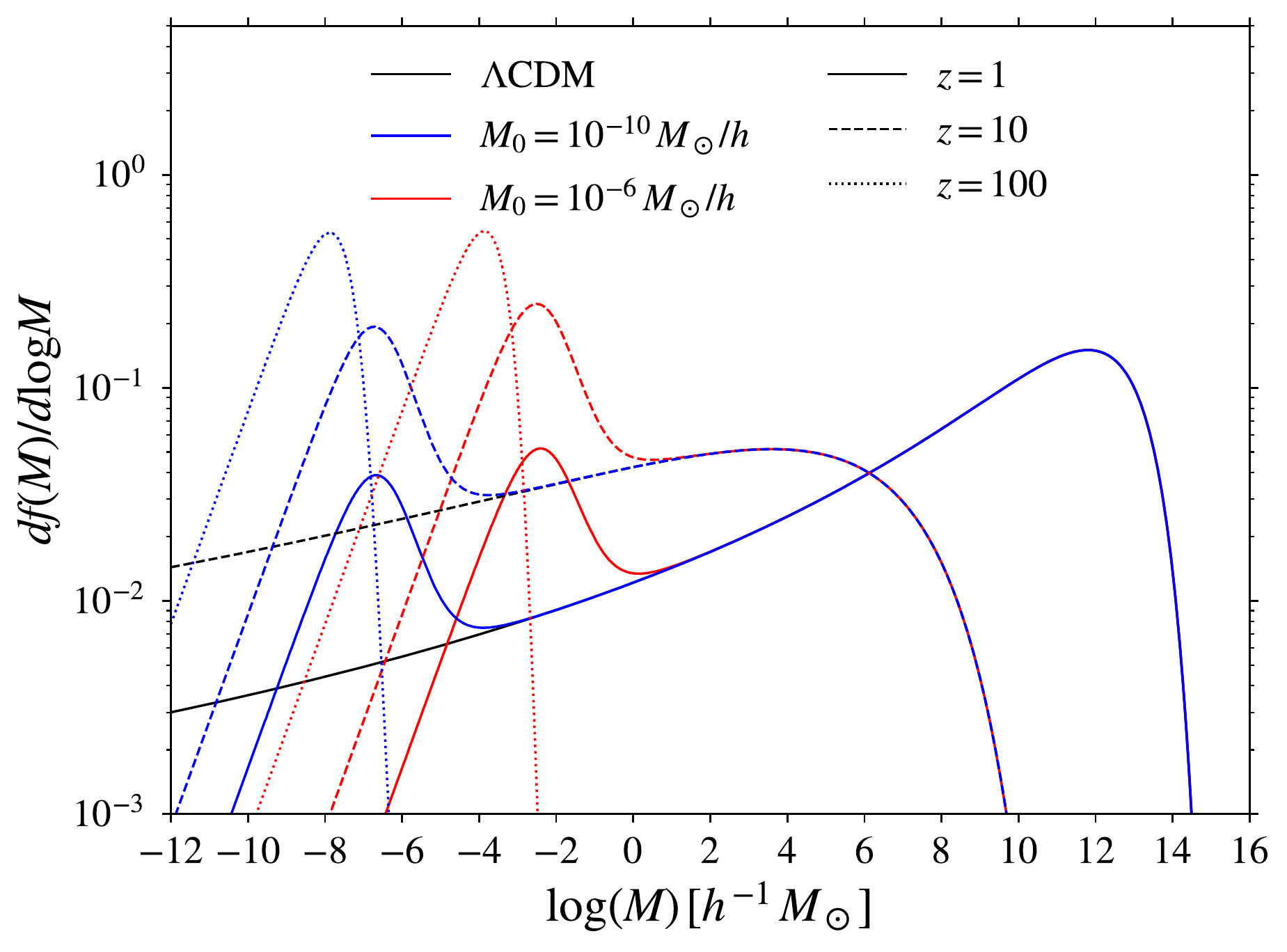}
  \caption{Differential mass fraction of {\it isolated} collapsed halos, $\rmd f(M)/\rmd \log M$, computed from the Press-Schechter formalism. We compare the standard CDM cosmology (black) and the axion cosmology with two different values for the $M_0$ parameter, $10^{-10}\,h^{-1}\,M_\odot$ (blue) and $10^{-6}\,h^{-1}\,M_\odot$ (red), at three different redshifts: $z=1$ (solid), $z=10$ (dashed), and $z=100$ (dotted).} 
  \label{fig:dfdlogM}
\end{figure}

In \reffig{dfdlogM}, we adopt the PS formalism to compute $\rmd f/\rmd\log M$ at various redshifts. We consider different values of $M_0$ for the axion cosmology and compare the results to those of the standard cosmology. As an example, for $M_0 = 10^{-10}\,h^{-1}\,M_\odot$, by $z=100$ virtually all DM is locked up in collapsed axion minihalos of masses $M < 10^{-6}\,h^{-1}\,M_\odot$, well before any galactic scale halos form. As hierarchical assembly proceeds, small minihalos gradually assemble into larger halos. By $z=10$, about $50\%$ of DM resides in isolated minihalos of masses $M < 10^{-4}\,h^{-1}\,M_\odot$, and a substantial fraction has already assembled into the first dwarf-galaxy halos from the nonlinear collapse of adiabatic fluctuations. By $z=1$, around the time when the most massive galaxy clusters are forming, isolated minihalos of masses $M < 10^{-4}\,h^{-1}\,M_\odot$ still contribute about $10\%$ of all DM. This provides a lower limit on the mass budget of minihalos of $M < 10^{-4}\,h^{-1}\,M_\odot$ in the Universe, because we have excluded subhalos of similar masses whose contribution depends on the substructure survival fraction and is not accounted for by the PS formalism.

\reffig{dfdlogM} shows that structure formation in axion cosmology differs substantially from that in the standard cosmology because the collapse times of the very low-mass halos shift to earlier epochs. In the standard cosmology, practically no DM structures have collapsed on any scales by $z=100$. The lowest mass halos form only at $z \sim 20$, at an epoch not well separated from the formation times of the first halos of dwarf galaxy scale. For the axion cosmology with $M_0=10^{-10}\, h^{-1}\, M_\odot$, isolated halos with $M \lesssim 10^{-4}\,h^{-1}\,M_\odot$ account for $\sim 30\%$ of all DM. By $z=1$, this fraction drops to $\sim 10\%$ as more of them merge into larger halos. On the other hand, the DM mass fraction (including substructure) residing in large halos converges between the standard cosmology and the axion cosmology. The PS calculation indicates that in axion cosmology, the mass fraction in tiny isolated minihalos with $M < 10^{-4}\,h^{-1}\,M_\odot$ is much higher than that in standard cosmology. Moreover, the low-mass axion minihalos are much denser, so they are much more likely to survive as substructure when they merge into larger halos. When the value of $M_0$ is increased, for example, to $M_0 = 10^{-6}\,h^{-1}\,M_\odot$, the characteristic minihalo masses shift to larger masses; the aforementioned conclusions remain qualitatively true after numbers are appropriately rescaled.

The PS calculation includes only isolated minihalos, whereas observations of caustic transiting stars behind lensing clusters are predominantly probing DM subhalos inside galaxy clusters rather than intervening field halos in intergalactic space (c.f. discussion in \refsec{powerspec2d}). Nevertheless, we make the following argument to translate the above results to the {\it subhalo} mass fraction inside a galaxy cluster, up to caveats regarding minihalos surviving dynamic disruption in the intracluster environment. Most of the mass of a galaxy cluster accretes at $z \sim 1$ from the intergalactic medium, where the abundance of small DM halos shortly before the accretion to the cluster is well described by the PS mass function. Being much denser than the cluster halo, the axion minihalos are unlikely to be disrupted once they merge into the cluster. The fractional mass function of minihalos inside the cluster halo can therefore be approximately described by \refeq{dfdlnM}. This approximation is still a lower limit to the number of minihalos, because the PS formalism does not include minihalos that collapse and survive as subhalos in intermediate mass halos, and are incorporated into the cluster when the intermediate mass object merges into the cluster.

Using this justification, we shall use \refeq{dfdlnM} to estimate the power spectrum of the surface density projected along the line of sight through a cluster halo in \refsec{powerspec2d}. The various theoretical uncertainties regarding dynamical disruption of minihalos and the multiple levels of subhalos inside subhalos will affect our result only by a moderate factor reflecting the fraction of DM in axion minihalos.

\section{Cumulative effect of minihalos along a line of sight}
\label{sec:powerspec2d}

The main goal of this paper is to analyse the gravitational lensing impact of axion minihalos in the line of sight to a source observed at high magnification. As mentioned, axion minihalos are difficult to detect with gravitational lensing because of their low surface density. However, small surface density fluctuations $\Delta \Sigma$ produce large effects near critical curves, when the lensing magnification is as high as $\sim \Sigma_{\rm crit}/\Delta\Sigma$.

Near the critical curve of a galaxy cluster, the total cluster surface density is close to $\Sigma_{\rm crit}$, whereas the surface density of an individual minihalo is much lower. This implies that if, as argued above, axion minihalos contain a substantial fraction of all the DM, their area covering factor is much larger than unity. Therefore, axion minihalos must move through each other on many high-speed encounters as they orbit inside the cluster halo. At the same time, surface density fluctuations near a lensing critical curve result from the superposition of many axion minihalos, and can therefore be treated as a Gaussian random field, with all the information being included in the power spectrum. We evaluate the surface density power spectrum in this section.

\subsection{Surface density power spectrum}

A useful starting point of the calculation is to consider that all the relevant density fluctuations are contained in a statistically homogeneous slab of thickness $L$, which is much larger than any transverse scales of interest. As shown in \refapp{proj}, the power spectrum of the surface density field $\Sigma$ is in this case related to that of the volumetric density field $\rho$ by $P_\Sigma(q_\perp) = P_\rho(q_\perp)\,L$, where $q_\perp$ is the two-dimensional Fourier wave number conjugate to the transverse length scale in proper units. This also assumes that the power spectrum is not too blue, obeying $\rmd\ln P_\rho(q)/\rmd\ln q < 1$, where $q$ denotes the three-dimensional Fourier wave number.

If all mass is locked into subhalos of some mass $M$ and number density per unit $\ln M$ given by $n(M) = (\bar\rho/M)\, \rmd f(M)/ \rmd\ln M$, the volumetric density power spectrum is \citep[see][]{Cooray:2002dia},
\ba
\label{eq:Prhoq}
P_\rho(q) = \int\,\rmd \ln M\, n(M)\,\left| \tilde{\rho}^{(h)}\left(q; M \right) \right|^2 = \overline{\rho}\,\int\,\frac{\rmd M}{M^2}\,\frac{\rmd f(M)}{\rmd \ln M}\,\left| \tilde{\rho}^{(h)}\left(q; M \right) \right|^2.
\ea
where $\overline{\rho}$ is the mean density, and
\ba
\tilde{\rho}^{(h)}(q; M) := 4\pi\,\int^{\infty}_0\, r^2\,\rmd r\,\frac{\sin\left(q\,r\right)}{q\,r}\,\rho^{(h)}\left( r;\,M \right).
\ea
is the Fourier transform of the halo density profile (assumed to be spherical symmetric) at halo mass $M$. In principle, the subhalo mass distribution does not have to be the same as the field distribution in \refeq{dfdlnM} considered in the previous section, but by the argument put forward in \refsec{massfun} we will use the approximation here that they are the same.

We now consider a long slab along the line of sight through the cluster halo. The total length of the slab is of the order of the cluster virialization scale $\sim$Mpc. The scale of the relevant transverse modes is comparable to the radii of the axion minihalos, which is orders of magnitude smaller. Since the coarse-grained density profile varies substantially from the inner region to the outer region of the cluster halo, it is invalid to assume that the volumetric density field is statistically homogeneous when deriving the power spectrum of the surface density field across the slab. Therefore, we divide the slab along the line of sight into many sub-slabs. Each sub-slab is sufficiently thin for the density field to have approximately homogeneous statistics, but is thick enough for the derivation of \refapp{proj} to be valid. The surface density field through the entire slab has a power spectrum equal to the uncorrelated sum of the contributions from all sub-slabs. 

We make another simplifying premise that all sub-slabs share the same {\it fractional} mass distribution as a function of halo mass, $\rmd f/\rmd \ln M$. Then the halo mass function is linearly proportional to the local mean density in each sub-slab. According to \refeq{Prhoq}, the power spectrum for the surface density field is given by a summation over all sub-slabs, or an integral along the line of sight in the limit of a large number of sub-slabs,
\ba
P_\Sigma(q_\perp) & = & \left( \int\,\frac{\rmd M}{M^2}\,\frac{\rmd f(M)}{\rmd \ln M}\,\left| \tilde{\rho}^{(h)}\left(q_\perp; M \right) \right|^2 \right)\,\left( \int\,\rmd L\,\overline{\rho}(L) \right) \nonumber\\
& = & \overline{\Sigma}\,\left( \int\,\frac{\rmd M}{M^2}\,\frac{\rmd f(M)}{\rmd \ln M}\,\left| \tilde{\rho}^{(h)}\left(q_\perp; M \right) \right|^2 \right),
\ea
where $\int\,\rmd L\,\left( \cdots \right)$ is the line-of-sight integral. The fractional surface overdensity, defined as $\deltwod := \Sigma/\barSigma$, has a power spectrum
\ba
P_{\deltwod}(q_\perp) = \frac{1}{\barSigma}\,\left( \int\,\frac{\rmd M}{M^2}\,\frac{\rmd f(M)}{\rmd \ln M}\,\left| \tilde{\rho}^{(h)}\left(q_\perp; M \right) \right|^2 \right).
\ea
Under our assumptions, only the mean surface density $\barSigma$ is needed for computing $P_{\deltwod}(q_\perp)$, provided that the (uniform) fractional mass distribution $\rmd f/\rmd \ln M$ and the halo density profiles are known.

Apart from minihalos residing in the cluster halo, intervening minihalos free floating in the intergalactic space are also expected. Indeed, in substructure tests utilizing galaxy scale lensing, intervening field halos integrated over a cosmological distance can be equally important or even dominant compared to subhalos~\citep{Despali:2017ksx}. By comparison, the amount of substructure through a rich galaxy cluster is substantially larger. Assuming similar mass fractions of minihalos inside the cluster halo and in the intergalactic space, the relative importance of the two contributions can be inferred from comparing the surface density through the cluster $\overline{\Sigma}_{\rm cl}$ and the integrated DM surface density everywhere else along the line of sight $\overline{\Sigma}_{\rm los}$. For the two known highly magnified stars behind MACS J1149~\citep{Kelly:2017fps} and behind MACS J0416~\citep{Chen:2019ncy, Kaurov:2019alr}, we find $\overline{\Sigma}_{\rm los} \approx 0.1$--$0.2\,\overline{\Sigma}_{\rm cl}$, which indicates that intergalactic minihalos should be subdominant.

It is convenient to measure the surface mass density in units of $\Sigma_{\rm crit}$, which gives the lensing convergence $\kappa = \Sigma/\Sigma_{\rm crit}$. Neglecting intergalactic minihalos, the lensing convergence has a power spectrum,
\ba
\label{eq:Pkappa}
P_\kappa(q_\perp) = \frac{\barSigma_{\rm cl}}{\Sigma^2_c}\, \int\,\frac{\rmd M}{M^2}\,\frac{\rmd f(M)}{\rmd \ln M}\,\left| \tilde{\rho}^{(h)}\left(q_\perp; M \right) \right|^2.
\ea
We emphasize that this calculation is conservative because it only counts minihalos formed in isolation prior to cluster formation and excludes minihalos who themselves orbit intermediate mass subhalos (i.e. subhalos of subhalos). 

\begin{figure}
  \includegraphics[width=0.49\columnwidth]{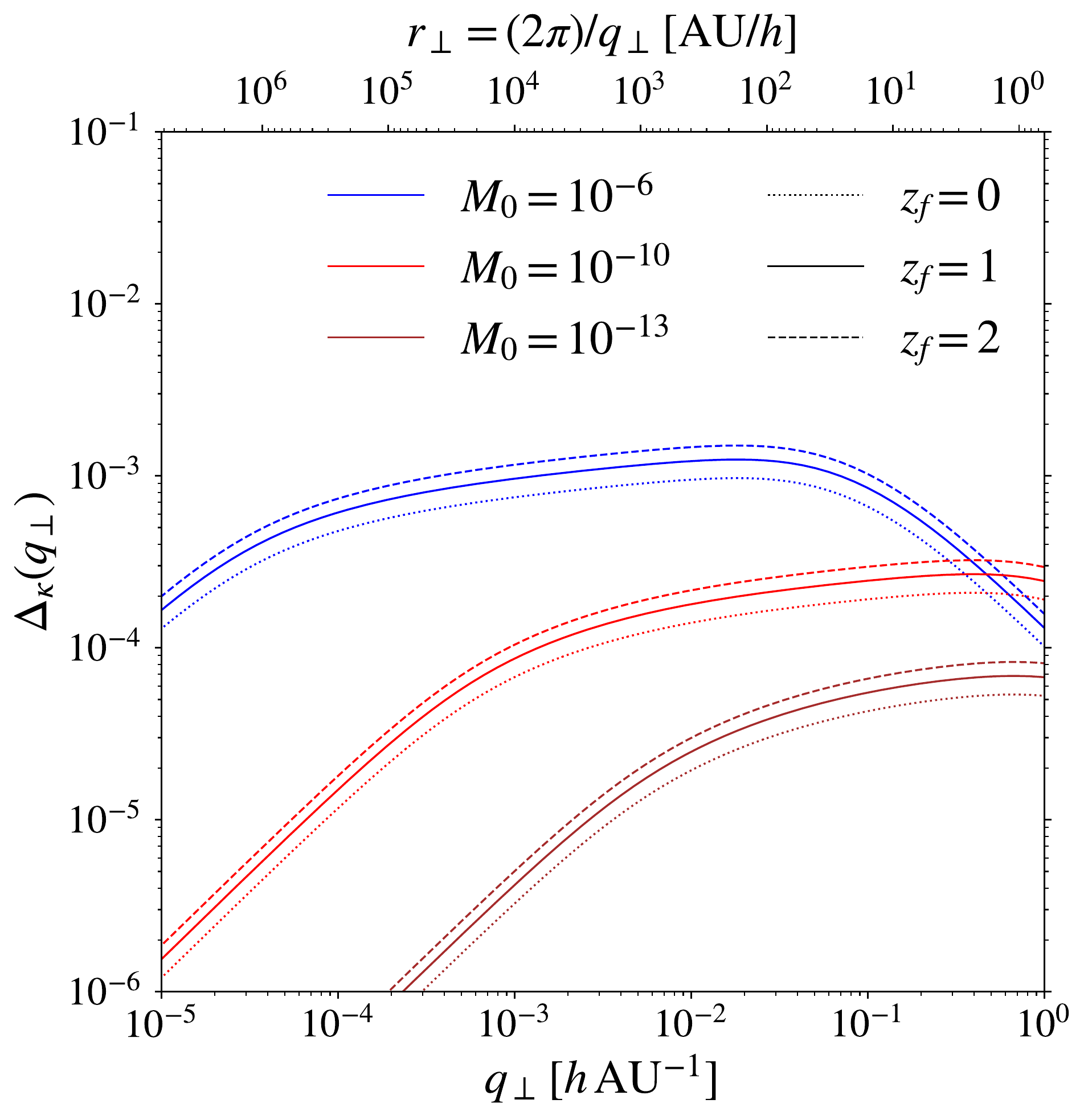}
  \includegraphics[width=0.51\columnwidth]{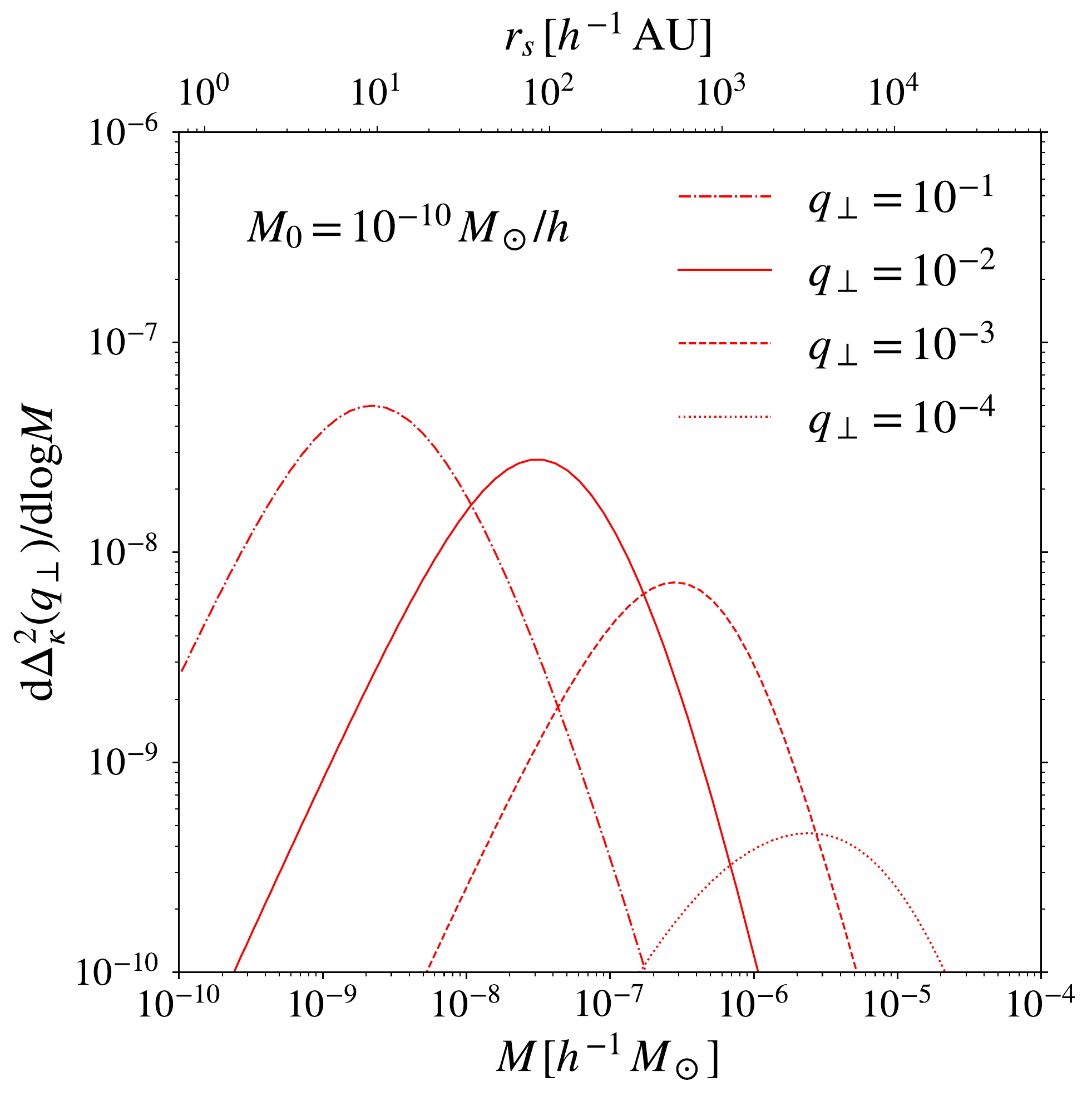}
  \caption{ The case of the line of sight to the caustic-straddling lensed galaxy behind the lensing cluster MACS J1149. {\it Left panel}: Characteristic fluctuation $\Delta_\kappa(q_\perp)$ in the convergence $\kappa$ as a function of the projected Fourier wave number $q_\perp$. When evaluating the PS mass function $\rmd f/\rmd \ln M$ (c.f. \refeq{dfdlnM}), we set the redshift $z_f=0$ (dotted), $1$ (solid), and $2$ (dashed). We plot curves for three different values for the axion cosmology characteristic mass scale (in units of $M_\odot/h$) $M_0=10^{-6}$ (blue), $10^{-10}$ (red), and $10^{-13}$ (brown). The top axis indicates the corresponding projected proper length scale $r_\perp = 2\pi/q_\perp$. {\it Right panel}: Differential contribution to the variance of the convergence from axion minihalos of different masses at given $q_\perp$. We set $M_0 = 10^{-10}\,M_\odot/h$, and show curves for $q_\perp=10^{-1}$ (dash-dotted), $10^{-2}$ (solid), $10^{-3}$ (dashed) and $10^{-4}$ (dotted), all in units of $h/{\rm AU}$. The top axis indicates the corresponding minihalo scale radius $r_s$. Calculations exclude subhalos inside subhalos.} 
  \label{fig:Deltaka}
\end{figure}

In \reffig{Deltaka}, we present a calculation of the convergence power spectrum for the specific case of the line of sight toward the highly magnified star LS1 behind MACS J1149. In the calculation, we assume that axion minihalos inside the cluster halo obey the PS mass function \refeq{dfdlnM}, but neglect additional contributions to the clumpy surface density due to those minihalos having their own satellite minihalos. 

In the left panel of \reffig{Deltaka}, we compute $\Delta_\kappa(q_\perp):=\left[ q^2_\perp\,P_\kappa(q_\perp)/(2\pi) \right]^{1/2}$, which quantifies the characteristic fluctuation in the lensing convergence $\kappa$ as a function of the transverse length scale. For $M_0 = 10^{-10}\,M_\odot/h$, $\kappa$ can fluctuate at the level of $\Delta_\kappa \sim 10^{-4}$ on scales $r_\perp = (2\pi)/q_\perp = 10$--$10^4\,{\rm AU}/h$, which, as we will explain in \refsec{microlensing}, are probed by the light curves of the highly magnified stars during events of microlensing peaks. The right panel of \reffig{Deltaka} shows that at a fixed wave number $q_\perp$, the convergence fluctuation originates mainly from minihalos with a typical scale radius $r_s \simeq r_\perp = (2\pi)/q_\perp$. 

For an increased value for the parameter $M_0=10^{-6}\,M_\odot/h$, axion minihalos of $M  \lesssim 10^{-2}\,M_\odot/h$ form earlier and thus are more compact. As a result, the typical convergence fluctuation on those scales can increase to $\Delta_\kappa \sim 10^{-3}$. If instead $M_0$ is several orders of magnitudes smaller, axion minihalos in the mass range of our interest become less dense, and the resultant convergence fluctuation is reduced. 

The left panel of \reffig{Deltaka} indicates convergence fluctuations at the level of $\Delta_\kappa \sim 10^{-4}$--$10^{-3}$ for a broad range $M_0=10^{-13}$--$10^{-6}\,M_\odot/h$, which translates into the mass scale of the earliest gravitational collapse $\sim 0.01\,M_0 =10^{-15}$--$10^{-8}\,M_\odot/h$. The results are not very sensitive to the choice of the cluster formation redshift $z_f$ at which the PS mass function is evaluated.

\subsection{Area covering factor}

Assuming a uniform $\rmd f/\rmd \ln M$ throughout the cluster halo, we can calculate the optical depth to intersecting a minihalo of mass $M$ within the scale radius $r_s(M)$. Taking a geometric cross section $\pi\,r^2_s$, the differential optical depth is
\ba
\label{eq:dtaudlnM}
\frac{\rmd \tau(M)}{\rmd \ln M} = \frac{\pi\,r^2_s(M)\,\barSigma_{\rm cl}}{M}\,\frac{\rmd f(M)}{\rmd \ln M}.
\ea
This can be interpreted as the area covering factor.

\begin{figure}
 \centering
  \includegraphics[width=0.49\columnwidth]{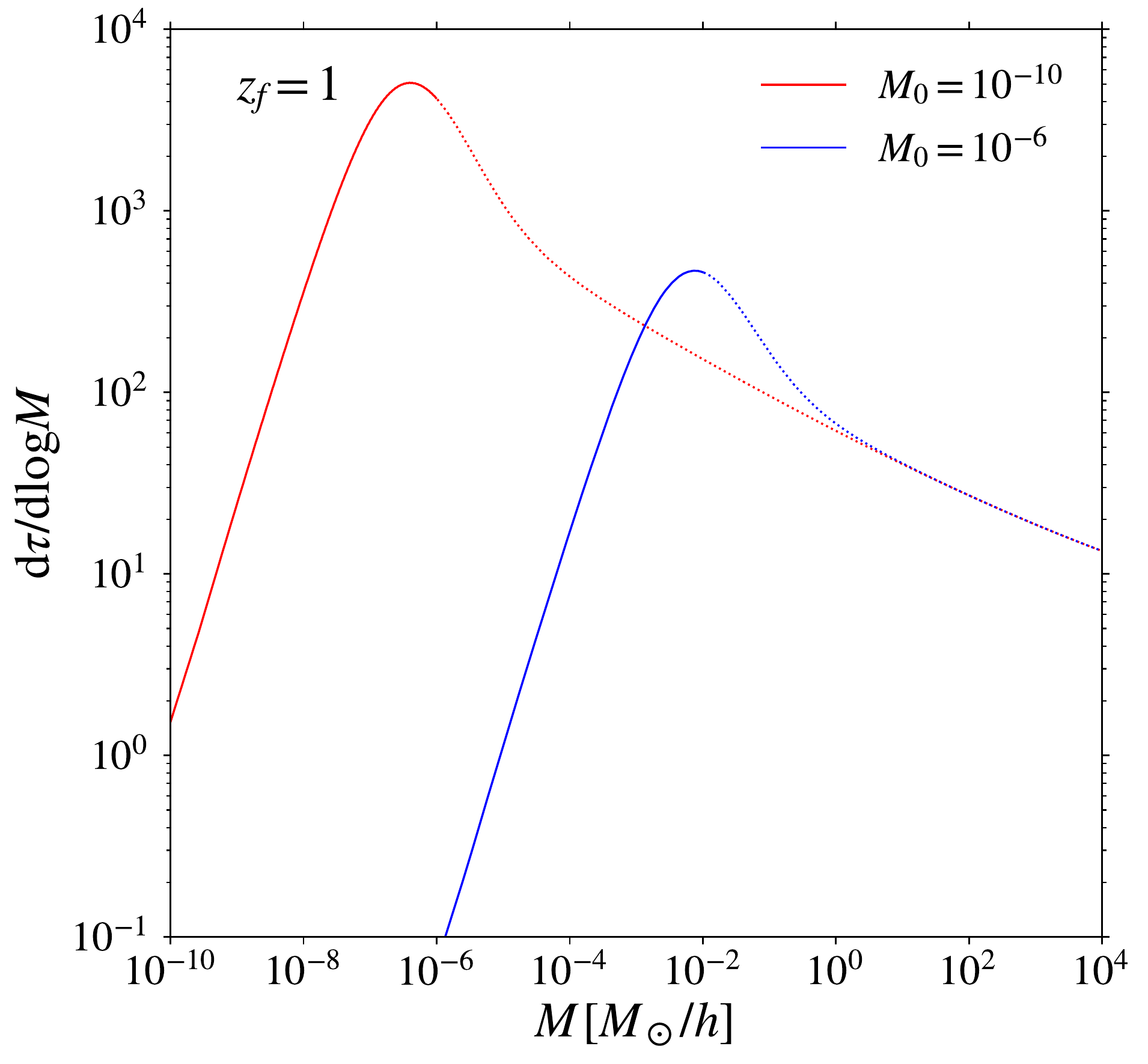}
  \caption{Differential area covering fraction $\rmd\tau/\rmd \log M$ for axion minihalos of a range of masses (\refeq{dtaudlnM}). We consider two values for the axion cosmology characteristic mass scale $M_0 = 10^{-10}\,M_\odot/h$ (red) and $10^{-6}\,M_\odot/h$ (blue). For each curve, the left portion (solid) corresponds to halos that form from isocurvature density fluctuations, and the right portion (dotted) corresponds to halos that form from adiabatic density fluctuations. } 
  \label{fig:coverfactor}
\end{figure}

\reffig{coverfactor} plots this optical depth as a function of the halo mass, for $M_0 = 10^{-10}$ and $10^{-6}\,M_\odot/h$, respectively. Taking $M_0 = 10^{-10}\,M_\odot/h$ as the example, we find that minihalos formed from isocurvature density fluctuations $M \sim 10^{-10}$--$10^{-6}\,M_\odot/h$ generally have $\rmd\tau/\rmd \log M \gg 1$. Within this mass range, as increasingly massive (and hence physically bigger) minihalos are considered, a single line of sight traversing the entire cluster halo intersects an increasingly larger number of minihalos. Since these minihalos are the major contributors to convergence fluctuations on scales of $\sim 10$--$10^4\,{\rm AU}/h$, the convergence field, by the Central Limit Theorem, is locally well described by a Gaussian random field with an isotropic power spectrum \refeq{Pkappa}. The super-critical area covering fraction for the axion minihalos is in sharp contrast to their volume occupation fraction, which is tiny because of the characteristic density of minihalos being many orders of magnitude higher than that of the cluster halo.

As for the more massive halos $M \gtrsim 10^{-6}\,M_\odot/h$ formed from the adiabatic density fluctuations, the area covering factor starts to decrease. That part of the curve is shown as dotted, because the area covering factor is certainly overestimated toward larger $M$. Applying the PS mass function $\rmd f/\rmd \ln M$ to intracluster substructures severely overestimates the abundance of those halos inside the cluster halo as simulations suggest that many of them do not survive disruption. This is because the density hierarchy between those halos and the cluster halo is substantially narrower, and because across several decades of masses those halos collapse at similar redshifts and therefore all have similar densities~\citep{diemand2006early}. It should be clear to readers that those halos are not the focus of this study.

In the case of $M_0 = 10^{-6}\,M_\odot/h$, we reach a similar conclusion for minihalos of masses $M \sim 10^{-4}$--$10^{-2}\,M_\odot/h$, although the overlap in projection for the least massive minihalos ($M \lesssim 10^{-4}\,M_\odot$) are not high enough to render the convergence field completely Gaussian. Finally, we note that our evaluation of \refeq{dtaudlnM} has left out minihalos inside subhalos or minihalos into higher levels of substructure hierarchy. Including those will further Gaussianize the convergence field and increase its power spectrum.

\section{Dynamical disruption of minihalos inside clusters}
\label{sec:disruption}

We now discuss the dynamical processes that determine the survival rate of axion minihalos orbiting within larger CDM halos. We consider minihalos of mass $m_1$ and radius $r_1$ orbiting within a parent halo of mass $M$ and radius $R$. Those account for a fraction of the total host halo mass $\rmd f_{\rm sub}(m_1;M)/\rmd\ln m_1$. We distinguish this from the PS mass fraction $\rmd f(m_1)/\rmd \ln m_1$, which neither accounts for dynamic destruction of subhalos, nor for the fact that minihalos of mass $m_1$ exist at multiple levels of substructure hierarchy (subhalos of subhalos, and so on), all of which participate in dynamical processes.

We discuss three important dynamical processes that determine the destruction rate of minihalos:
 
\begin{enumerate}
     \item Spiraling to the host halo center by dynamical friction, and tidal disruption at the high central halo mass density (which may be enhanced by baryonic processes in galactic centers such as disk and bulge formation).
     \item Dynamical heating during high-speed encounters with other minihalos. 
     \item Dynamical heating during high-speed encounters with baryonic structures, including stars and molecular clouds.
\end{enumerate}
 
We will show that among these destruction mechanisms, the most important is the second one if only DM is present, and the third one after stars have formed. Our estimates suggest that, over the age of the Universe, minihalos of our concern are unlikely to be disrupted by a significant fraction in a typical galaxy cluster when stars in member galaxies are taken into account. Some useful order-of-magnitude formulae are collected in \refapp{dynamic}. 

\subsection{Dynamical friction and tidal disruption near the halo center}
\label{subsec:disdf}

For a subhalo of mass $m_1$ orbiting inside a parent halo of mass $M$, the timescale for orbital decay due to dynamical friction is~\citep{binney2011galactic}
 \begin{equation}\label{eq:tdf}
     t_{\rm df} \approx t_{\rm orb}\, \left( \frac{M}{m_1} \right)\,\left(\ln\frac{M}{m_1}\right)^{-1},
 \end{equation}
After a few $t_{\rm df}$'s, the subhalo can spiral into the very inner part of the parent halo where they are destroyed by the tidal force of the dense central mass concentration. 

\refeq{tdf} suggests that this mechanism is effective on minihalos if the parent halo is a larger minihalo so that $M/m_1$ is not too large. The orbital timescale is evaluated in \refeq{torbapp}. For example, a minihalo with $m_1 = 10^{-8}\,M_\odot/h$ can sink to the center of a parent minihalo with $M = 10^{-6}\,M_\odot/h$ within $\sim 3\,$Gyr. Precise understanding of this hierarchical assembly and destruction of minihalos from white noise initial density fluctuations will require a dedicated study that employs numerical simulations.

However, many minihalos should evade dynamic friction constraints. First, the PS analysis in \refsec{massfun} has shown that a significant fraction of minihalos can be directly bound into either the cluster halo or its galactic-scale subhalos without first being assembled into a moderately larger parent halo or a series of slightly larger parent halos. For these minihalos, the humongous $M/m_1$ ratio ensures immunity to dynamic friction.

Second, even if a minihalo is first bound into a moderately larger parent halo and starts to spiral in, the parent may be subsequently disrupted by an even larger halo. In this case, the increased $M/m_1$ ratio allows the minihalo to survive for a longer time. This is especially relevant if the first parent halo is one on the portion of the colored curve in \reffig{rs_vs_rhos} that levels off. In this case, this halo is very susceptible to disruption because subsequent hierarchical mergers take place quickly and involve small density differences. If the original minihalo is quickly ``liberated'' into orbiting a very large and low density parent halo, then $t_{\rm df}$ becomes certainly too long compared to the age of the Universe.
 
A subhalo can be destroyed even without spiraling in, simply by being randomly placed into a highly radial orbit toward the very inner part of a larger halo. This can lead to a high-speed encounter with either the central DM density cusp, or possibly with other highly condensed form of matter and compact objects. We will see in the following subsections that these fine-tuned encounters are not the most important processes compared to encounters with other minihalos and ordinary stars.
 
 \subsection{High-speed encounters with other subhalos}
\label{subsec:dishe}

The large optical depth calculated for \refeq{dtaudlnM} of \refsec{powerspec2d} and the fact that over the cluster's age minihalos should have completed many orbits inside the cluster halo imply that direct encounters between minihalos are frequent.

We now consider the dynamical heating experienced by a subhalo of mass $m_1$ and internal velocity dispersion $\sigma_1$, when passing close to or moving through another subhalo of mass $m_2$ due to a random encounter. We focus on high-speed encounters for which the typical encounter relativity velocity\footnote{Our chosen notation $\sigma$ is motivated because the typical encounter velocity is set by the internal velocity dispersion of the host halo. This should not be confused with the peak height $\sigma(M, t)$ of the PS formalism.} $\sigma$ is much greater than other velocity scales~\citep{aguilar1985tidal, binney2011galactic}.

We first consider the case $m_2\simeq m_1$. Then, when the two subhalos fly through each other at an impact parameter $b$ smaller than the subhalo scale radii $r_1$, the velocity perturbation induced on any particle of the subhalos is $\Delta v \simeq \sigma^2_1/\sigma$, which is a small fraction of $\sigma_1$. This is because the tidal perturbation during the encounter is comparable to the orbital acceleration within one halo, but the duration of the encounter is shorter than the orbital time by a factor $\sigma_1/\sigma$. The fractional amount by which the internal subhalo energy varies owing to dynamical heating at each encounter with $b < r_1$ is therefore
\begin{equation}\label{eq:frace}
    \frac{\Delta E_1}{E_1} \sim \frac{\sigma_1^2}{\sigma^2}.
\end{equation}
Dynamical heating scales as the second power of $\Delta v/\sigma_1$ if $\Delta v/\sigma_1 \ll 1$. At the level of individual bound particles of halo $m_1$, the leading order fractional change in the particle's energy is linearly proportional to $\Delta v/\sigma_1$, but can be either positive or negative depending on the orbital phase during the encounter. After the halo relaxes back to equilibrium, the average energy injection scales as $(\Delta v/\sigma_1)^2$.

For unequal encounters, the minimum impact parameter is the larger $r_s$ of the two halos. Larger impact parameters are less important, because the induced velocity perturbations cale as $b^{-2}$, and the internal dynamical heating as $b^{-4}$, whereas the rate of encounters within $b$ increases only as $b^2$.

In the case $m_1 < m_2$, an impact parameter $b$ smaller than the scale radius $r_2$ of the perturber halo $m_2$ leads to an encounter with the inner density cusp. For an NFW slope of the halo inner density profile, the cusp mass scales as $b^2$. The velocity perturbation $\Delta v$ is therefore comparable to that if $b \simeq r_2$, but the probability for a smaller impact parameter is suppressed as $\propto b^2$. This implies that inner density cusps are not the dominant sources of dynamic heating, and it is justified to take the larger $r_s$ of the two halos to be the minimum impact parameter.

At the minimum impact parameter, the induced velocity perturbation is maximized for $m_2 \simeq m_1$. Encounters with a less massive perturber, with $m_2 < m_1$, has $\Delta v \propto m_2$. For encounters with a higher mass halo with $m_2 > m_1$, we focus on minihalos that collapse at high redshifts from the white noise isocurvature density fluctuations, for which the scale radius goes as $r_2 \propto m^{5/6}_2$. In this case, the tidal acceleration scales as $m_2\,r^{-3}_2 \propto m_2^{-3/2}$, and the encounter time as $r_2$, so $\Delta v \propto m_2^{-2/3}$. Since even in the worst case $m_2 \simeq m_1$ we have $\Delta v /\sigma_1 \ll 1$, particle ejections by a single tidal shock are unlikely, and hence the quadratic scaling \refeq{frace} is justified.

We now show that the {\it cumulative} heating from many encounters is still dominated by nearly equal encounters with $m_2 \simeq m_1$. For the case $m_2 < m_1$, dynamical heating per encounter scales has $m^2_2$, while the encounter rate will scale as $m_2^{-1}\,\rmd f_{\rm sub}(m_2;\,M)/\rmd \ln m_2$. If the PS mass function \refeq{dfdlnM} is applicable to subhalos, $\rmd f_{\rm sub}(m_2;\,M)/\rmd \ln m_2 \propto m_2^{1/2}$ for minihalos and the encounter rate goes as $m^{-1/2}_2$. In principle, subhalos of subhalos or multiple levels of subhalos also contribute to the disrupting tidal field, which are not included in \refeq{dfdlnM}. Without a detailed calculation, we argue that including those $\rmd f_{\rm sub}(m_2;\,M)/\rmd \ln m_2$ should depend logarithmically on $m_2$ because minihalos are not easy to disrupt through hierarchical merging, and accordingly the encounter rate goes as $m^{-1}_2$. In either case, the increase in the encounter rate does not compensate for the decreased heating per encounter. For the case $m_2 > m_1$, dynamical heating per encounter decreases as $m_2^{-4/3}$, but the encounter rate increases only as $r_2^2\,m^{-1}_2\,\rmd f_{\rm sub}/\rmd \ln m_2 \propto m_2^{2/3}\,\rmd f_{\rm sub}/\rmd \ln m_2$, again insufficient to compensate for the decreased heating per encounter. The above analysis should be valid unless the more massive minihalos have somehow increased their density by dissipative processes above the value reached through virialization at the collapse epoch.
 
Encounters between halos of comparable masses with impact parameters comparable to their common scale radii have a rate inside the cluster halo, 
\ba
    \Gamma_{\rm enc}(m_1; M) \sim \frac{1}{m_1}\,\frac{\rmd f_{\rm sub}(m_1; M)}{\rmd\ln m_1}\,\frac{R}{t_{\rm orb}(M)}\, \frac{r^2_1\,M}{R^3} ~,
\ea
where $R$ is the scale radius of the cluster halo. Given the fractional heating per encounter \refeq{frace}, the timescale of dynamic disruption is
\ba
    t_{\rm dispt}(m_1; M) \sim t_{\rm orb}(M)\,\frac{R}{r_1}\,\left( \frac{\rmd f_{\rm sub}(m_1; M)}{\rmd\ln m_1} \right)^{-1},
\ea
where we have assumed virial equilibrium for both the minihalo and the host halo $R/r_1=(M/m_1)\,(\sigma_1/\sigma)^2$. This result can also be expressed in terms of $t_{\rm df}$ of \refeq{tdf},
\ba
    t_{\rm dispt}(m_1;M) \sim t_{\rm df}(m_1;\,M)\,\left( \frac{\sigma_1}{\sigma} \right)^2\,\left( \frac{\rmd f_{\rm sub}(m_1; M)}{\rmd\ln m_1} \right)^{-1}\,\ln \frac{M}{m_1} ~.
\ea
Since the velocity dispersion for minihalos is much smaller than that of the host halo (see \refeq{sigmaapp}), we have $t_{\rm dispt} < t_{\rm df}$. The dominant process for minihalo destruction is therefore dynamical heating by high-speed encounters among comparable minihalos, where they repetitively go through each other before they are disrupted by gradual heating and expansion. Numerically, the disruption timescale is (where we have inserted the relation $G\,M/R \sim \sigma^2$ for the host halo by the virial theorem)
\ba
t_{\rm dispt} & \approx & 2 \times 10^8\,{\rm Gyr}\,\left( \frac{\rmd f_{\rm sub}(m_1; M)}{\rmd\ln m_1} \right)^{-1}\,\left( \frac{M}{10^{15}\,M_\odot/h} \right)^2\,\left( \frac{\sigma}{10^3\,{\rm km/s}} \right)^{-5}\nonumber\\
&& \times \frac{\nu}{\nu_{\rm med}} \, \left( \frac{\Omega_a}{0.3} \right)^{1/3}\,\left( \frac{1+z_{\rm eq}}{3400} \right)\,\left( \frac{M_0}{10^{-10}\,M_\odot/h} \right)^{1/2}\,\left( \frac{m_1}{10^{-6}\,M_\odot/h} \right)^{-5/6}.
\ea
Even for the largest possible $\rmd f_{\rm sub}(m_1; M)/\rmd\ln m_1$, this is much longer than the age of the Universe within cluster-size or galaxy-size hosts, and even true for minihalos orbiting small CDM halos that form from the adiabatic density fluctuations, say one with $M = 10^6\,M_\odot/h$ and $\sigma= 5\,{\rm km/s}$. 

\subsection{High-speed encounters with stars}
\label{subsec:dishe}

After the gas in halos condenses into stars and other compact objects, the potential for tidal disruption of minihalos is greatly enhanced. The reason for perturber minihalos of similar masses ($m_2\simeq m_1$) dominating the destruction rate by high-speed encounters rather than perturber minihalos of larger masses has to do with the fact that the more massive minihalos are also less dense, with $\rho_s \propto m_1^{-3/2}$ in the regime that minihalos collapse and assemble from the white-noise initial isocurvature density fluctuations. By contrast, for a compact perturber $m_2$, the minimum impact parameter is fixed by the size of the minihalo $m_1$, then the amount of dynamical heating per encounter increases with the mass of the perturbing object as $m_2^2$. If the number density of perturbers only decreases approximately as $m_2^{-1}$, dynamical heating should be dominated by the most massive perturbers.

Tidal interactions between generic small halos and passing stars have been extensively studied~\citep{goerdt2007survival, green2007mini, 2007MNRAS.375.1146A, Zhao:2005mb}. Here we present a simple order-of-magnitude estimate applied to axion minihalos. A minihalo of mass $m_1$ and radius $r_1$ encountering a star with mass $m_\star$, at an impact parameter $b$ and a relative velocity on the order of the host halo's internal velocity dispersion $\sigma$ is subject to a tidal acceleration $\sim G\,m_\star r_1/b^3$ during the encounter time $\sim b/\sigma$, and suffers an impulsive velocity perturbation,
\begin{equation}
    \frac{\Delta v}{\sigma_1} \simeq \frac{G\,m_\star\,r_1}{b^2\,\sigma\,\sigma_1}.
\end{equation}
Unlike mutual encounters between minihalos, the velocity perturbation induced by a stellar perturber can be large enough that minihalo particles are unbound after a single encounter. This happens for $\Delta v/\sigma_1 \simeq 1$, which sets a minimum impact parameter $b_{\rm min}$ for the minihalo to survive a complete disruption~\citep{goerdt2007survival, schneider2010impact, Tinyakov:2015cgg},
\ba
    \frac{b_{\rm min}}{r_1} & \simeq & \left( \frac{G\,m_\star}{\sigma^2\,r_1}\,\frac{m_\star}{m_1} \right)^{1/4} 
    \approx 0.7\,\left( \frac{m_\star}{M_\odot} \right)^{1/2}\,\left( \frac{\sigma}{10^3\,{\rm km/s}} \right)^{-1/2} \nonumber\\
    && \times \left( \frac{\nu}{\nu_{\rm med}} \right)^{1/4}\, \left( \frac{\Omega_a}{0.3} \right)^{1/12}\,\left( \frac{1+z_{\rm eq}}{3400} \right)^{1/4}\,\left( \frac{M_0}{10^{-10}\,M_\odot/h} \right)^{1/8}\,\left( \frac{m_1}{10^{-6}\,M_\odot/h} \right)^{-11/24}.
\ea
For a range of minihalo parameters, $b_{\rm min}$ is comparable or larger than the minihalo scale radius $r_1$. Provided that the overall tidal disruption rate is dominated by encounters with small impact parameters, we therefore conclude that a minihalo is disrupted whenever it passes within $b_{\rm min}$ from a perturber star.

Let $\bar{n}_\star$ be the {\it mean} stellar number density within the host halo (including the stellar populations in all member galaxies). The stellar disruption timescale for minihalos orbiting within the host halo can be estimated as
\ba
\label{eq:tdisptstar}
t_{\rm dispt, \star} & \sim & \left( \pi\,b^2_{\rm min}\,\bar{n}_\star\,\sigma \right)^{-1} = \frac{m_1^{1/2}}{\pi\,r_1^{3/2}\,G^{1/2}\,m_\star\,\bar{n}_\star} \approx 50\,{\rm Gyr}\, \left( \frac{m_\star}{M_\odot} \right)^{-1}\,\left( \frac{\bar{n}_\star}{10^5\,{\rm kpc}^{-3}} \right)^{-1} \nonumber\\
&& \times \left( \frac{\nu}{\nu_{\rm med}} \right)^{3/2}\, \left( \frac{\Omega_a}{0.3} \right)^{1/2}\,\left( \frac{1+z_{\rm eq}}{3400} \right)^{3/2}\,\left( \frac{M_0}{10^{-10}\,M_\odot/h} \right)^{3/4}\,\left( \frac{m_1}{10^{-6}\,M_\odot/h} \right)^{-3/4}.
\ea
Here the fiducial value $\bar{n}_\star= 10^5\,{\rm kpc}^{-3}$ roughly corresponds to enclosing $\sim 10^{13}\,M_\odot$ of stars within a sphere of $\sim 300\,{\rm kpc}$. Note that the internal velocity dispersion of the host halo $\sigma$ cancels out in this calculation. \refeq{tdisptstar} suggests that the most massive minihalos are the most vulnerable because they are the puffiest. For minihalos in the most interesting mass range concerning this work $m_1 < 10^{-6}\,h^{-1}\,M_\odot$, and in axion cosmology scenarios with $M_0 > 10^{-10}\,M_\odot/h$, stellar disruption is not significant throughout the current age of the Universe. More massive minihalos however are expected to have been destroyed by passing stars in an intracluster environment.

By chance minihalos can pass the central regions of larger halos at high speed on radial orbits. Apart from stars, supermassive black holes residing at the center of member galaxies are probably the most disruptive perturbers, for which the previous analysis remains applicable. \refeq{tdisptstar} implies that the disruption timescale is inversely proportional to the average mass density of the perturbers within the cluster halo. Despite being much more destructive individually, supermassive black holes still make a smaller contribution to the total mass budget than the stars do and hence are not the major sources of disruption. 

Finally, we note that even if all minihalos orbiting within the intracluster space are dynamically destroyed, we still expect a remaining contribution from an intergalactic population of minihalos. Those minihalos are not subject to dynamic disruption from encounters with stars. Some of those are isolated, while others are bound to larger intergalactic halos. The latter are susceptible to additional dynamic disruption from mutual encounters inside the parent halo.

\section{Impact on microlensing light curves of highly magnified stars}
\label{sec:microlensing}

Despite several theoretical uncertainties, our analyses in the previous sections lead us to the following physical picture if a Peccei-Quinn phase transition in the early Universe induces isocurvature density fluctuations on small scales:

\begin{itemize}

    \item Minihalos below the mass scale of planets form by gravitational instability starting at the epoch of radiation-matter equality, with characteristic densities much higher than in the standard CDM cosmology.
    
    \item Plenty of these minihalos should survive over the age of the Universe as subhalos orbiting inside the halos of galaxy clusters.
    
    \item The high area covering factor of minihalos implies that a line of sight traversing a cluster lens near a typical lensing critical curve (with an impact parameter of tens of kiloparsecs) probes nearly Gaussian random fluctuations in the lensing convergence $\kappa$ (and hence also in the shear). For the mass scale parameter $M_0 \sim 10^{-13}$--$10^{-6}\,M_\odot/h$, which corresponds to the first gravitationally collapsed minihalos having massses $\sim 10^{-15}$--$10^{-8}\,M_\odot/h$, the level of fluctuations is $\Delta\kappa \gtrsim 10^{-4}$--$10^{-3}$, on scales $\sim 10$--$10^4\,{\rm AU}/h$.
    
\end{itemize}

In most applications of gravitational lensing, convergence fluctuations of $\sim 10^{-4}$--$10^{-3}$ on the lens plane are too small to lead to any interesting observational consequences. However, caustic transiting stars behind a galaxy cluster lens provide extreme situations in which these minuscule small-scale fluctuations in the convergence can leave observable imprints.

\subsection{Detectable scales of surface density fluctuations under microlensing}

We consider the effect of convergence and shear irregularities on the total flux of a point source. The flux magnification of one image equals the inverse determinant of the lensing Jacobian matrix at the image position, whose matrix elements are expressed in terms of the convergence and shear. When an image is highly magnified ($|\mu| \gg 1$), the corresponding Jacobian matrix is fine-tuned to a level $\sim 1/|\mu|$ to be nearly degenerate. Any irregularity in the convergence and/or shear at a level $\sim 1/|\mu|$ substantially perturbs the image flux.

The detected highly magnified stars behind MACS J1149~\citep{Kelly:2017fps} and behind MACS J0416~\citep{Chen:2019ncy, Kaurov:2019alr} are observed thanks to magnification factors of hundreds to thousands. If the projected mass distribution in the cluster lens were smooth, the magnification would simply vary as the inverse of the angular separation from the image to the critical curve, and could reach values above a million when a luminous star crosses the cluster caustic.
However, intracluster stars introduce
intermittent microlensing flux variations and micro-caustic crossings at which a pair of micro-images dominate the flux. This makes the highly magnified stars more easily identifiable from their variability, even though the maximum magnifications reached are reduced to $\sim 10^4$~\citep{Venumadhav:2017pps, Diego:2017drh, Oguri:2017ock}. Flux variations of up to a factor of ten have been observed from microlensing events in the fields of MACS J1149~\citep{Kelly:2017fps} and of MACS J0416~\citep{Chen:2019ncy}. During micro-caustic crossings, with typical durations of days to weeks, the star flux becomes susceptible to even minuscule nonsmoothness of $\Delta \kappa \sim 10^{-4}$ in the lens surface density.

During a micro caustic transit event, the two dominant micro-images are highly elongated to a length $\approx 5000\,{\rm AU}\,(|\mu|/10^4)\,(R_\star/100\,R_\odot)$. Surface density irregularities are detectable down to these scales, which are interestingly comparable to the axion minihalo sizes we have discussed.

\subsection{Micro-Fold Model}

We now demonstrate how minuscule surface density fluctuations imprint irregularities in microlensing light curves. \refapp{geoopt} shows that geometric optics is applicable to our problem.

For a concrete case study, we consider the highly magnified star LS1 detected in a lensed galaxy at $z_S =1.49$ behind the cluster lens MACS J1149 at $z_L=0.544$~\citep{Kelly:2017fps}. This sets the angular diameter distances to the source $D_S = 1.79\,$Gpc, to the lens $D_L = 1.35\,$Gpc, and from the lens to the source $D_{LS}= 0.95\,$Gpc. Two resolved macro-images of LS1 are observed at a separation of $0.13''$ from the cluster critical curve, symmetrically positioned around a point where the macro lens model \texttt{GLAFIC}~\citep{kawamata2016precise} predicts a surface mass density corresponding to a convergence $\kappa_0 = 0.83$, and a magnification for each macro-image of $\sim 300$.

The presence of microlenses implies that each macro-image actually consists of several unresolved micro-images. During a transit of a fold caustic (the most common caustic type), a pair of micro-images usually become much brighter than the combined flux from all the rest (which is in any case nearly constant). This justifies considering only the two dominant micro-images; instead of simulating a network of microlensing caustics induced by many intracluster stars, constructing a simple fold model for the micro-caustic crossing suffices. The fold model is described by two parameters: the local convergence $\tilde \kappa_0$ and an eigenvalue gradient vector $\boldsymbol{\tilde d}$ at the point between the two images along the micro-critical curve. The total smooth convergence $\kappa_0$ is split into a smooth contribution $\tilde \kappa_0$, including the DM and any diffuse baryonic component, and the average contribution from stellar objects, which is $\kappa_\star\approx 0.005$ near the LS1 images in MACS J1149~\citep{Oguri:2017ock}. We use $\tilde \kappa_0 = \kappa_0 - \kappa_\star \simeq 0.83$ as a good approximation.

The vector $\boldsymbol{\tilde d}$ is perpendicular to the micro-caustic and, in the fold approximation, is related to the peak magnification $\mu_{\rm pk}$ at the time when the source stellar disk grazes the micro-caustic on the source plane:
\ba
\label{eq:mupk}
\mu_{\rm pk} & = & \frac{1}{2\,|1-\tilde\kappa_0|}\,\left( \frac{D_S}{|\boldsymbol{\tilde d}|\,|\sin\tilde \alpha|\,R_S} \right)^{1/2} \nonumber\\
& \approx & 4\times 10^4\,\left(\frac{|1-\tilde \kappa_0|}{0.17}\right)^{-1}\,\left( \frac{|\boldsymbol{\tilde d}|\,|\sin\tilde\alpha|}{100\,{\rm arcsec}^{-1}} \right)^{-1/2}\,\left(\frac{R_S}{10\,R_\odot}\right)^{-1/2}\,\left(\frac{D_S}{1\,{\rm Gpc}}\right)^{1/2},
\ea
where $\tilde \alpha$ is the angle between the micro-critical curve and the degenerate direction of the local Jacobian matrix, and the stellar radius is $R_S$. This fold approximation is accurate when the highly magnified images are much closer to the micro-critical curve than the typical angular scale of variation of the vector $\boldsymbol{\tilde d}$.

The presence of microlenses generates a region around the macro-critical curve of the cluster lens where the micro-critical curves interact and join together, forming a network of width $\sim \kappa_\star |\boldsymbol{\tilde d}|^{-1}$ \citep{Venumadhav:2017pps}. Within this network, the scale of variation of $\boldsymbol{\tilde d}$ is about the mean separation between microlenses, $\sim \theta_\star/\kappa_\star^{1/2}$, where $\theta_\star$ is the Einstein angular radius of each individual microlens of mass $M_\star$, $\theta_\star=(4\,G\,M_\star/D_{\rm eff}\,c^2)^{1/2}$, with $D_{\rm eff}=D_L\, D_S/D_{LS}$. At the same time, the typical magnitude of $\boldsymbol{\tilde d}$ in the corrugated network is
\ba
\label{eq:btd}
|\boldsymbol{\tilde d}| \sim \kappa^{3/2}_\star/\theta_\star \sim 350\,{\rm arcsec}^{-1}\,\left(\frac{\kappa_\star}{0.01}\right)^{3/2}\,\left( \frac{M_\star}{1\,M_\odot} \right)^{1/2}\,\left(\frac{D_{\rm eff}}{1\,{\rm Gpc}}\right)^{-1/2}.
\ea
For demonstrating typical cases, we shall set in the next subsection $|\boldsymbol{\tilde d}|=\kappa^{3/2}_\star/\theta_\star$, with fiducial values for $\kappa_\star$ and $M_\star$ as in the above equation. We shall focus on situations of high magnification, with $\mu_{\rm pk}\gtrsim 10^3$--$10^4$, which requires the lensed star to be close enough to a micro-caustic. In fact, the typical average magnification of an image at an angular separation
$\theta_\star/\kappa^{1/2}_\star$ from a micro-critical curve is $\mu \sim |1-\tilde\kappa_0|^{-1}\,(|\boldsymbol{\tilde d}|\,|\sin\tilde \alpha|\,\theta_\star\, \kappa_\star^{-1/2})^{-1/2} \sim |1-\tilde \kappa_0|^{-1}\,\kappa^{-1/2}_\star \ll 10^3$--$10^4$, for $\kappa_\star \sim 10^{-2}$. For the high magnification of our interest, the two dominant micro images have a much smaller angular separation than $\theta_\star/\kappa^{1/2}_\star$, justifying our use of a fold model locally. 

The timescale of caustic transients depends on an effective source velocity $v_t$. Roughly, this is the relative transverse velocity between the source star and the lens cluster \citep[defined in Eq.(12) of][]{Venumadhav:2017pps}, which is constant. Although intracluster stellar micro-lenses and DM substructure have random velocities of $\sim 1000\,{\rm km/s}$ within the host cluster, their impact is dwarfed by the even larger apparent velocity of the micro images on the image plane owing to the large magnification factor. Therefore, for the numerical simulations in the next subsection, the assumptions of a stationary micro-fold caustic model and time independent realizations of axion minihalos along the line of sight are justified.

\subsection{Numerical Examples}

\begin{figure}
 \centering
  \includegraphics[width=\columnwidth]{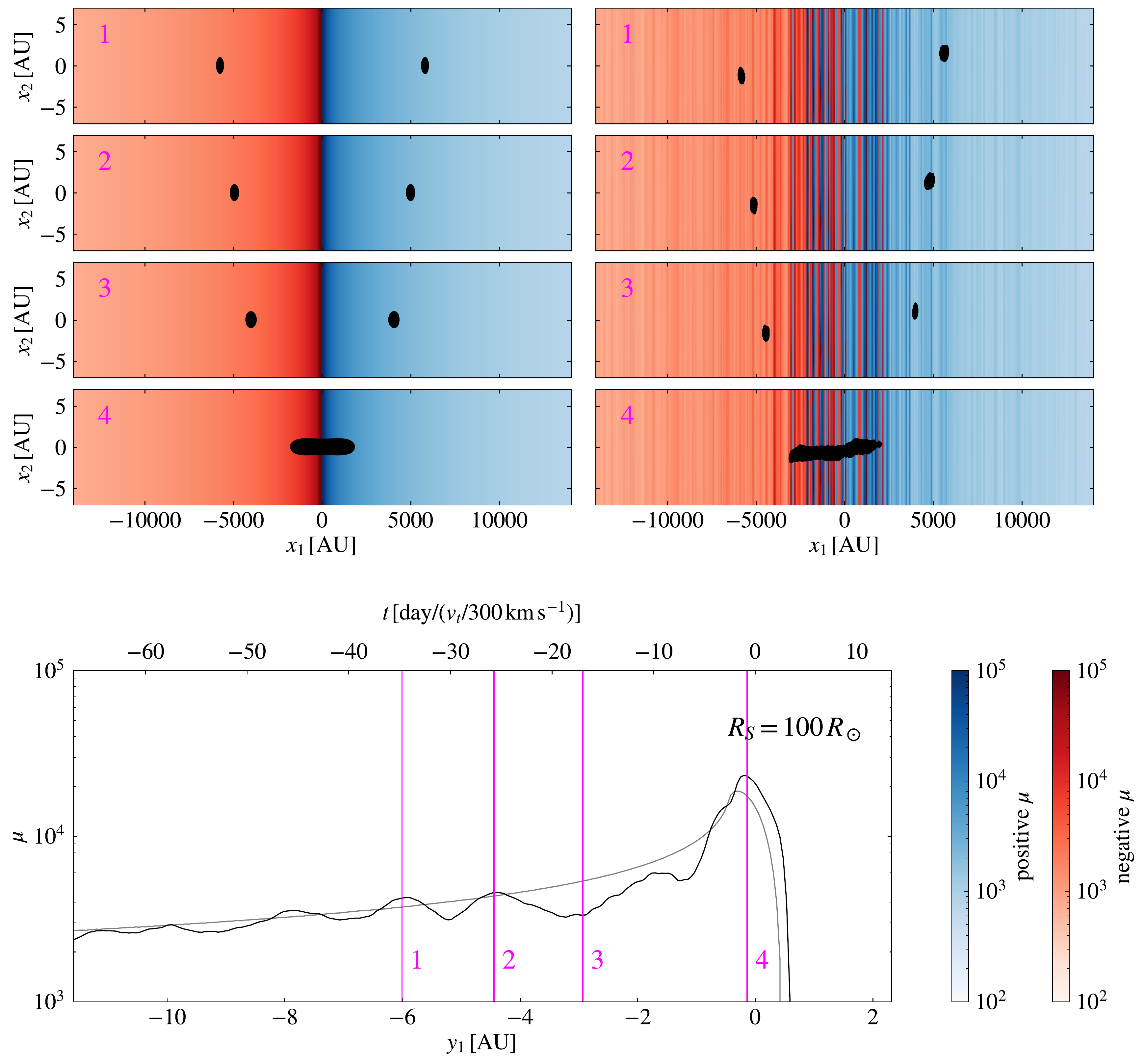}
  \caption{Example of a microlensing peak event for a highly magnified supergiant star of $R_S = 100\,R_\odot$, and macro lensing parameters measured for MACS J1149 LS1. For axion cosmology, we set $M_0 = 10^{-10}\,M_\odot/h$. We set $\tilde \alpha=\pi/2$ and $|\boldsymbol{\tilde d}|=\kappa^{3/2}_\star/\theta_\star$, with $\kappa_\star=0.005$ and $\theta_\star$ computed for a micro-lens mass $M_\star = 0.3\,M_\odot$. {\it Top four rows:} Dominant micro-image pair in the image plane (in proper length units) and magnification pattern including its sign, with (right column) and without (left column) small-scale surface density fluctuations due to axion minihalos. Coordinates $x_1$ and $x_2$ (shown on drastically different scales!) are parallel and perpendicular, respectively, to the degeneracy direction of the local micro-fold model. Each row corresponds to a numbered epoch. Color scales for magnification are shown to the right of the bottom panel. {\it Bottom row}: Total flux magnification versus the source one-dimensional position $y_1$ (in proper units) in the source plane, and the variability timescale converted from an effective source velocity $v_t$, with (black curve) and without (grey curve) surface density fluctuations due to minihalos. The coordinate $y_1$ measures the position perpendicular to the micro-caustic. Only contributions from the two dominant micro-images are included. The four numbered epochs examined in the top rows are marked by magenta lines.} 
  \label{fig:lc_example_RS100}
\end{figure}

\begin{figure}
 \centering
  \includegraphics[width=\columnwidth]{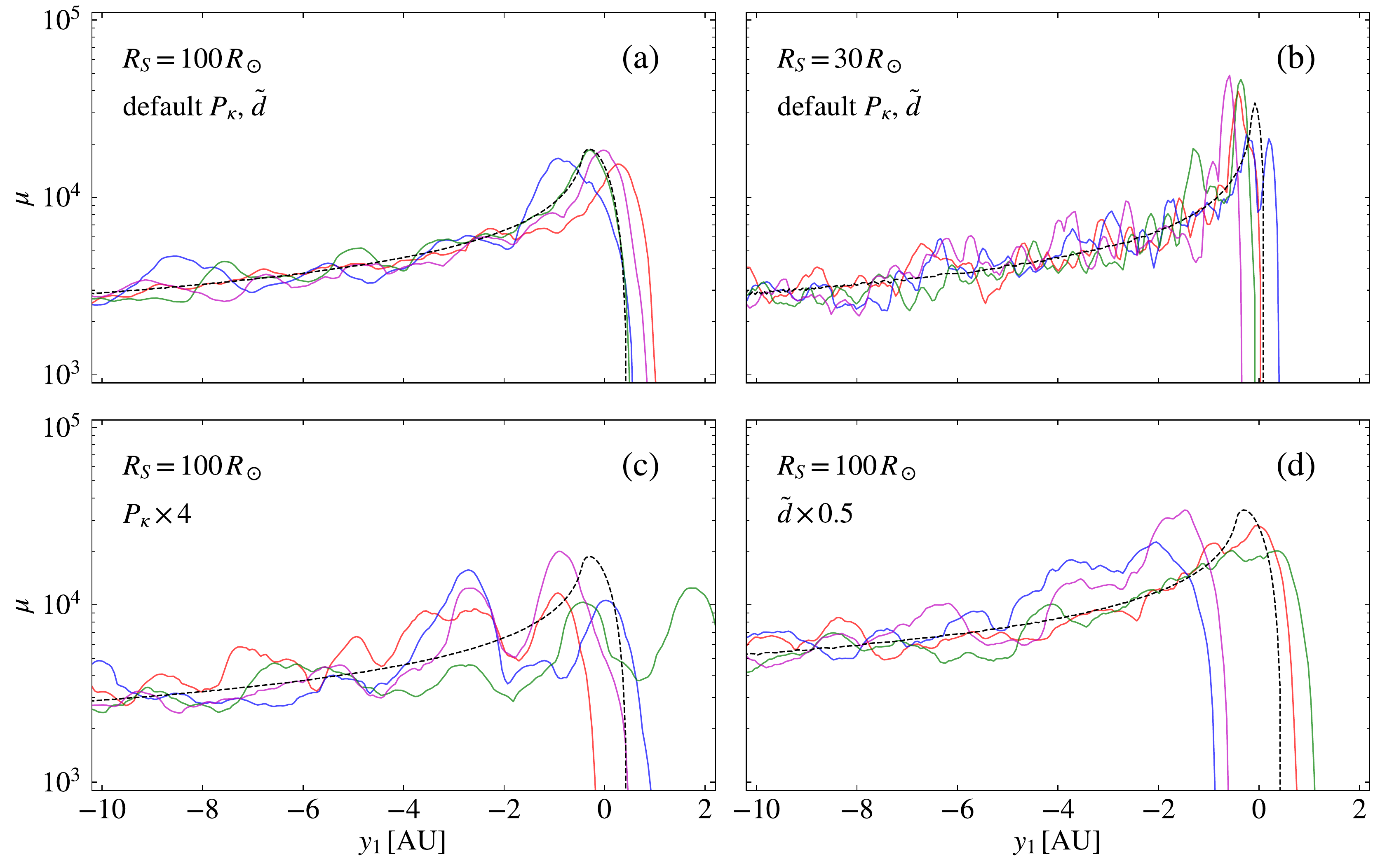}
  \caption{Perturbed light curves (colored curves) compared to smooth light curve (dashed black curve) around the time of a microlensing peak event. Each panel shows four random realizations of convergence fluctuations (one color for each). (a) Default case as in \reffig{lc_example_RS100}. (b) A more compact source star with $R_S = 30\,R_\odot$. (c) Power spectrum $P_\kappa$ enhanced by a factor of four. (d) $\tilde d=|\boldsymbol{\tilde d}|$ decreased by a factor of two.} 
  \label{fig:lc_example_dependence}
\end{figure}

The analytical estimates in \refeq{mupk} and \refeq{btd} suggest that microlensing peak magnifications can reach $\sim 10^3$--$10^4$ for reasonable lens and source parameters. We employ a numerical simulation to verify that fractional fluctuations $\sim 10^{-4}$--$10^{-3}$ in the lens surface density on appropriate scales are sufficient to imprint observable effects in the total flux.

On top of the micro-fold model, we add a spatially varying convergence perturbation $\Delta\kappa$ due to minihalos. Without having to generate individual minihalos, we are justified to model $\Delta\kappa$ as a Gaussian random field as shown in \refsec{powerspec2d}. We generate random realizations of $\Delta\kappa$ according to the homogeneous and isotropic power spectrum in \refeq{Pkappa}, converting the linear Fourier wave vector to the angular Fourier wave vector $\bfell=D_L\,\bfq_\perp$. In the case of MACS J1149, the critical surface density is computed to be $\Sigma_{\rm crit} = 2.3\times 10^9\,M_\odot/{\rm kpc}^2$, and $\bar\Sigma_{\rm cl}=0.83\,\Sigma_{\rm crit}$. The perturbed deflection required for inverse ray-tracing can then be computed in the Fourier domain $\Delta\bfalp(\bfell) = (-2\,i\,\bfell/\ell^2)\,\Delta \kappa(\bfell)$.

In \reffig{lc_example_RS100} we present a numerical example of how minihalos collectively induce irregularities in the light curve during a micro caustic transit event, assuming parameters appropriate for the case of MACS J1149 LS1. In the absence of minihalos, the total flux rises smoothly as $|t-t_*|^{-1/2}$ after the two highly elongated micro images become dominant, peaks when the finite source effect kicks in, and then plummets when the two micro images merge and disappear. Minihalos cause sizable brightening and fading "bumps" in the light curve despite the minuscule surface density fluctuations they induce. As the microlensing peak is approached, these irregularities become more prominent. In this example, the irregularities occur on a timescale of several days for $v_t \sim 300\, {\rm km/s}$. Unlike the usual microlensing signature of compact lenses, these irregularities lack abrupt changes; due to caustic crossings, microlensing light curves usually show asymmetric peaks, which have a slowly varying wing followed by an abrupt cutoff, or the time reversed behavior.

In the perturbed light curve the time of micro image merger shifts relative to that in the unperturbed one. This is due to coherent deflection perturbation generated by long wavelength modes. Causing merely a uniform remapping of the image plane coordinates, this bears no observable significance.

As visualized in \reffig{lc_example_RS100}, we observe that the perturbing effect is one dimensional in nature: only non-smoothness along the degenerate direction of the micro-fold matters as the two micro-images are highly elongated along this direction.

The perturbed light curve varies depending on a number of factors, as we show in \reffig{lc_example_dependence}. Firstly, the length of the elongated micro-image scales with the source size, which smooths out irregularities below certain scales. For a smaller source star, therefore, irregularities are present on shorter timescales. 

Secondly, larger convergence fluctuations will result in larger magnitude irregularities. This can be the case if more minihalos survive dynamic disruption and are bound to intermediate sized subhalos, since the amount of minihalos contributing to surface density clumpiness will then be more than what \refeq{Pkappa} predicts.

Moreover, susceptibility to lensing perturbations depends on the $\boldsymbol{\tilde d}$ vector of the micro-fold model. Although in \reffig{lc_example_RS100} we have set $|\boldsymbol{\tilde d}|=\kappa^{3/2}_\star/\theta_\star$, actually $|\boldsymbol{\tilde d}|$ varies among micro-caustic crossing events. At a fixed distance from the micro-caustic, a smaller $|\boldsymbol{\tilde d}|$ corresponds to a larger magnification of the two micro-images. This means that the flux is more prone to surface density perturbations and is subject to larger irregularities.

If the typical convergence fluctuation $\Delta\kappa$ greatly exceeds the inverse magnification factor, even a micro-image can be strongly perturbed and break into yet smaller images. For example, this is likely the case for the scenario with $M_0=10^{-6}\,M_\odot/h$ shown in the left panel of \reffig{Deltaka}. The underlying physical principle is analogous to the disruption of macro-images into micro-images due to a super-critical number density of intracluster microlenses~\citep{Venumadhav:2017pps, Diego:2017drh, Oguri:2017ock}, or due to a high abundance of CDM subhalos~\citep{Dai:2018mxx}. A detailed analysis of this regime of strong image disruption is beyond the scope of this work, but is certainly an important situation to consider. Excessively large surface density fluctuations on length scales being considered here should be constrained by microlensing observations because they would smooth out stellar microlensing peaks.

\section{Discussion}
\label{sec:discuss}

Minihalos are predicted to form in many cases when the dark matter is composed of axion particles that can solve the strong CP problem of QCD. So far, no viable method for detecting these axion minihalos through their gravitational lensing effect has been proposed. A generic problem to detect them is their low surface density. This paper proposes that axion minihalos can be detected during extreme magnification events when luminous stars cross micro-caustics produced by intracluster stars near the macro-critical curves of massive galaxy clusters. Substantial deviations from the predicted lightcurve of a fold micro-caustic crossing over timescales from hours to days will the signature of large numbers of axion minihalos which should superpose along a line of sight to produce nearly Gaussian fluctuations in the surface density. 

The predicted lightcurve will in general be affected by limb-darkening effects in the source star. These effects should nevertheless be accurately predictable once the spectral type of the source star is known. An irregular behavior of the flux might, however, arise for other physical reasons. We now discuss some of these other possible sources of deviations from a predicted lightcurve.

\subsection{Microlensing by Planet-Sized Lenses}
\label{sec:planet}

Planets or planetesimals inside the foreground galaxy cluster may imprint distortions onto stellar microlensing light curves. Inside the micro-critical curve network, and without sub-stellar lenses, the typical separation between a segment of micro-critical curve and nearby microlens stars is $\sim (\kappa_\star\,\Sigma_{\rm crit}/M_\star)^{-1/2}\approx 0.1\,{\rm pc}\,(\kappa_\star/0.01)^{-1/2}\,(D_{\rm eff}/1\,{\rm Gpc})^{1/2}\,(M_\star/0.3\,M_\odot)^{1/2}$. Bound planets orbiting their host stars much more closely than this do not induce independent critical curves. 

If any rogue or wide-orbit planets lie far (in projection) from any stellar critical curves, the minimal surface density for them to form a finer network of critical curves is $\kappa_{\rm ffp} \gtrsim 1/\bar{\mu}$, where $\bar{\mu}\sim 100$--$1000$ is the magnification factor of the image being observed. This would require a comparable amount of mass in rogue planets and stars, which is ruled out by Milky Way microlensing surveys implying $\kappa_{\rm ffp} \lesssim 10^{-3}\,\kappa_\star$~\citep{2011Natur.473..349S, 2017Natur.548..183M}. 

Free-floating planets which happen to lie in the vicinity of the stellar micro-critical curve can have greatly enhanced microlensing effects. If these planets have mass $M_{\rm ffp}$, corresponding to an angular Einstein scale $\theta_{\rm ffp}$, one requires $\kappa_{\rm ffp} \gtrsim (\theta_{\rm ffp}\,\tilde{d})^{2/3} \sim \kappa_\star\,(M_{\rm ffp}/M_\star)^{1/3}$ for breakup of stellar micro critical curves to be common~\citep{Venumadhav:2017pps}. This implies a much higher abundance of free-floating planets than main-sequence stars, which is unlikely.

If the stellar micro-critical curve is indeed substantially disrupted by sub-stellar lens objects \citep[see, e.g.,][for an analogous consideration of quasar microlensing]{2018ApJ...853L..27D}, finer variability structures are expected in the light curve near a stellar microlensing peak. However, in this case, signatures of planets or planetesimals would likely be sharp minor peaks rather than smoothly-behaved irregularities. 

\subsection{Blended Source Stars}
\label{sec:morestars}

The observed flux of the magnified star within its point spread function is likely to be contaminated by blended fainter stars. Although lensing of these fainter stars (or intrinsic variability if they are Cepheids) could contaminate the lightcurve, this is unlikely to happen because the highly magnified stars that are observed are usually among the most luminous in a galaxy, and they are sufficiently rare to make the superposition of two detectable lensed stars highly improbable. Contaminating microlensing is expected to produce sharp minor peaks in the light curve rather than the more smoothly-behaved irregularities from axion minihalos, as can be seen in \reffig{lc_example_RS100} and \reffig{lc_example_dependence}. If the contaminator is of a different stellar type, variability will be chromatic, which is not generally expected from lensing of a single star.

\subsection{Inhomogeneous gas}
\label{sec:diffsbaryon}

The dominant baryonic component inside a cluster halo is the ionized gas, which is shock heated to $T \sim 10^8\,$K and has a number density $n_e \sim 10^{-3}\,{\rm cm}^{-3}$. What is of our concern is the surface density fluctuations in the intracluster gas along the line of sight. However, what matters to the inquiry of this paper is inhomogeneities on minuscule scales $\sim 10$--$10^4\,$AU, which is orders of magnitude smaller than even the mean free path $\lambda_{\rm mfp} \sim 1\,$pc of the dilute gas. Although X-ray surface brightness measurements have clearly shown sizable turbulent density fluctuations $\Delta \delta_{\rm gas, 3d} \sim 10\,\%$ (fractional) on scales of $10$--$1000\,$kpc~\citep{Schuecker:2004hw, kawahara2008extracting, churazov2012x}, the characteristic amplitude of fluctuation is expected to decrease toward smaller scales following a Kolmogorov-like power-law $\Delta \delta_{\rm gas, 3d}(k) \propto k^{\alpha}$ with $\alpha \simeq -1/3$~\citep{Gaspari:2014vna}, and is likely to cut off well before scales of $1/k \sim 10$--$10^4\,$AU due to dissipation~\citep{Luan:2014iea}. For fixed volumetric density fluctuation, the surface density fluctuation is further suppressed by projection $\Delta \delta_{\rm gas, 2d}(q_\perp) \sim \Delta \delta_{\rm gas, 3d}(q_\perp)\,(q_\perp\,L)^{-1/2}$ (\refapp{proj}), where $L \sim 1\,$Mpc is the path length traversing the gas halo. Given the huge hierarchy between our projected scale of interest $1/q_\perp$ and $L$, roughly $q_\perp\,L \sim 10^7$--$10^{10}$, we hardly expect any detectable contribution to the lens surface density fluctuations from the hot diffuse gas. Cold and compact ISM gas structures of tiny sizes, if existent, may possibly perturb microlensing light curves, although their abundance and physical properties in the intracluster environment or in the surroundings of the BCG are poorly known. The lensing phenomenon we have been considering will probe or constrain any such small-scale baryonic structures. 

\section{Conclusion}
\label{sec:concl}

Recently discovered lensed individual stars behind cluster lenses have the largest magnification factors among all currently known gravitational lensing phenomena. Flux amplification by $\sim 10^3$--$10^4$ is expected when they transit microlensing caustics induced by intracluster stars. We have shown that, during these extreme magnification events, their lightcurves are susceptible to minuscule non-smoothness $\sim 10^{-3}$--$10^{-4}$ in the lens surface mass density, across very small projected scales $\sim 10$--$10^4\,{\rm AU}$, which are hardly accessible with other observational means.

Within the strongly motivated paradigm of axion DM particle, solar-system sized minihalos in the mass range $M \sim 10^{-10}$--$10^{-6}\,M_\odot$ are predicted to copiously orbit the DM halo of galaxy clusters. This prediction should be generic if the Peccei-Quinn symmetry breaking occurs after inflation. Our calculations have shown, for a promising range of axion mass, that a large number of minihalos along the line of sight are able to collectively induce detectable surface density fluctuations. This result applies to $M_0 = 10^{-13}$--$10^{-6}\,M_\odot/h$, translating into typical masses for the earliest gravitationally collapsed minihalos in the range $\sim 0.01\,M_0 \approx 10^{-15}$--$10^{-8}\,M_\odot$, an interval which includes the predicted parameters for the QCD axion. To our knowledge, monitoring microlensing variability of highly magnified stars is the first practical lensing-based method to probe axion minihalos.

Primordial black holes formed in the early Universe are another DM candidate~\citep{zel1967hypothesis, hawking1971gravitationally, carr1974black, carr2016primordial, garcia2017massive}. The mass budget of stellar mass black holes~\citep{bird2016did, clesse2017clustering, sasaki2016primordial} has been subject to stringent limits set by microlensing and dynamics. Recently, however, \cite{Montero-Camacho:2019jte} showed that the majority of DM may be made of black holes in the mass range $10^{-17}\,M_\odot \lesssim M_{\rm PBH} \lesssim 10^{-12}\,M_\odot$ without contradicting known astrophysical constraints. Such primordial black holes of asteroid masses may cluster into minihalos in a hierarchical fashion~\citep[see e.g.][]{Inman:2019wvr}, created from a white noise power spectrum in a similar way as axion minihalos, and they may produce similar lensing effects as the axion minihalos although with some detailed differences arising from dynamical relaxation and the discrete nature of black holes that may give rise to distinct observational signatures.

Accurate and high cadence light curves of highly magnified stars around the times of microlensing caustic transits have not been acquired so far. This will require dedicated monitoring using either space telescopes or large ground-based telescopes to reach the extremely faint fluxes at optical or near-infrared wavelengths, executed over days to weeks around the peak time with high cadence. In the case of stars lensed by MACS J1149 and MACS J0416, one or two microlensing peaks were mapped out only at crude levels. Since microlensing caustic transits occur randomly, an approximate time window bracketing the moment of flux culmination will need to be forecasted based on low cadence pre-monitoring that can detect a trend in which the star gradually brightens following a $f_\nu \propto |t-t_0|^{-1/2}$ law. Because any detected highly magnified star is generally expected to undergo intermittent micro-caustic transits for many years to follow, and the opportunity to discover DM minihalos or any other dark or baryonic small-scale lumpiness is unique and highly rewarding, there is a strong incentive to design and carry out these programs.

\begin{acknowledgments}
\acknowledgments

The authors would like to thank Matthew McQuinn, Scott Tremaine and Matias Zaldarriaga for insightful discussions, and Ken Van Tilburg for helping us clarify issues regarding the expected mass scale of QCD axion minihalos. We acknowledge the use of the publicly available Python package \texttt{Colossus} to carry out cosmological calculations~\citep{diemer2018colossus}. LD acknowledges the support from the Raymond and Beverly Sackler Foundation Fund. JM has been partly supported by Spanish fellowship PRX18/00444, and by the Corning Glass Works Foundation Fellowship Fund.

\end{acknowledgments}

\appendix

\section{Mass scale of first gravitationally collapsed minihalos}
\label{app:M0}

We evaluate the mass parameter $M_0$ defined in \refeq{M0} for the standard cosmological thermal history. In this Appendix, we adopt the natural units for which $c=\hbar=k_B=1$.

Let $T_0$ be the radiation temperature at time $t_0$ at the onset of axion field oscillation. The axion mean density at present is $\bar{\rho}_{a0}=(3\,H_0^2\,\Omega_a)/(8\,\pi\,G)$. In \refeq{M0}, we write $k_0=a(t_0)\,H(t_0)$. The scale factor at $t_0$ is
\ba
a(t_0) = \left( \frac{T_{\gamma 0}}{T_0} \right)\,\left[\left( 2 + 6\,\frac{7}{8}\,\left(\frac{T_{\nu 0}}{T_{\gamma 0}}\right)^3 \right)\frac{1}{g_{*s}}  \right]^{1/3},
\ea
where the temperature of the cosmic microwave background (CMB) $T_{\gamma 0}=2.725\,$K, the temperature of the relic neutrinos $T_{\nu 0}=(4/11)^{1/3}\,T_{\gamma 0}$, and $g_{*s}\approx 78$ is the effective number of relativistic degrees of freedom for entropy when the temperature is $\sim 1\,$GeV~\citep{husdal2016effective}. We calculate the Hubble parameter $H(t_0)$ using the Friedmann equation during the era of radiation domination $H(t_0)=[(8\,\pi\,G/3)\,(\pi^2/30)\,g_{*\epsilon}\,T^4_0]^{1/2}$, where $g_{*\epsilon}\approx 78$ is the effective number of relativistic degrees of freedom for energy density at a temperature $\sim 1\,$GeV~\citep{husdal2016effective}. Putting all pieces together, we find
\ba
M_0 = 10^{-9}\,M_\odot\,h^2\,\left(\frac{\Omega_a}{0.27}\right)\,\left(\frac{T_0}{1.3\,{\rm GeV}}\right)^{-3}\,\left(\frac{g_{*s}}{78}\right)\,\left(\frac{g_{*\epsilon}}{78}\right)^{-3/2}.
\ea
We note that our definition of $M_0$ is adopted from \cite{fairbairn2018structure}, which is a factor $4\,\pi^4/3$ larger than what is used in several other references~\citep[e.g.][]{Davidson:2016uok, hardy2017miniclusters}. Following the definition of the latter authors, the typical mass for the smallest minihalos that collapse the earliest from gravitational instability should be two orders of magnitude smaller than $M_0$. The exact number is not sharply defined because even these minihalos contain finer axion substructures as uncovered in numerical simulations.

\section{Linear growth of isocurvature modes}
\label{app:matter}

Under the assumption that the DM is made of axions, the isocurvature matter modes under consideration here are superhorizon modes at the onset of axion field oscillation, but enter the horizon during the era of radiaton domination. These perturbations do not grow during the radiation-dominated epoch. After the epoch of radiation-matter equality, they lock onto a growing mode that scales linearly with the scale factor.

To verify this picture, consider a toy universe composed of a radiation fluid and a matter fluid that does not interact with the radiation. We normalize the scale factor such that it is equal to unity $a_\star = 1$ when the mean radiation density equals the mean matter density $\rho_r(a_\star) = \rho_m(a_\star)$. Define a synchronous-comoving coordinate system in which the peculiar velocity of the matter fluid is always zero. We can derive the following set of equations~\citep{efstathiou1986isocurvature, Ma:1995ey}
\ba
h'' + \frac{2 + 3\,a}{2\,a\,(1+a)}\,h' & = & \frac{3}{a^2\,(1+a)}\,\left( 2\,\delta_r + a\,\delta_m \right), \\
\delta'_m & = & \frac12\,h', \\
\delta'_r & = & \frac23\,h' - \frac43\,\chi_r, \\
\chi'_r & = & \frac{\chi_r}{2\,a\,(1+a)} + \left( \frac{c\,k}{a_\star\,H_\star} \right)\,\frac{\delta_r}{2(1+a)}.
\ea
Here $a_\star$ and $H_\star = H(a_\star)$ are the scale factor and the Hubble parameter at the epoch of radiation-matter equality, respectively, $\delta_m$ and $\delta_r$ are matter and radiation overdensities, respectively, $h$ is the trace of the spatial metric perturbation in the synchronous gauge (to be distinguished from the dimensionless Hubble parameter $h$ elsewhere), and we have defined $\chi_r:=\theta_r/(a^2\,H)$ with $\theta_r$ being the velocity divergence of the radiation fluid. The notation $(\cdots)'$ represents the derivative with respect to the scale factor $a$.

We consider modes that enter the horizon well before the epoch of radiation-matter equality $c\,k/(a_\star\,H_\star) \gg 1$. The isocurvature initial condition is set at some initial time $a_i$ in the superhorizon regime $c\,k/(a_i\,H(a_i)) \ll 1$. At the initial time, the isocurvature initial condition requires $\delta_r(a_i) = - \delta_m(a_i)\,a_i$, $\chi_r(a_i) = h(a_i) = h'(a_i) = 0$.

In the limit of matter domination $a \rightarrow \infty$, the matter mode satisfies
\ba
\delta''_m + \frac{3}{2\,a}\,\delta'_m - \frac{3}{2\,a^2}\,\delta_m = 0,
\ea
which has a growing solution $\delta_m \propto a$.

By numerically solving the above set of linear equations, we find that deep in the era of matter domination $a \gg 1$ the solution for the matter perturbation is roughly
\ba
\delta_m(a) & = & C\,\frac{a}{a_\star}\,\delta_m(a_i),
\ea
where $C$ is an order-unity coefficient with a logarithmic dependence on $c\,k/(a_\star\,H_\star)$. The exact value of $C$ is cosmology dependent, but its value is not much different from unity.

We note that in principle the initial isocurvature overdensity can be much greater than unity. In this case, virialized structures can form well before the epoch of radiation-matter equality~\citep{kolb1994large, Enander:2017ogx}. A distribution of large initial overdensity patches were suggested by \cite{kolb1996femtolensing} according to numerical results. Microlensing constraints, e.g. suggested by \cite{Fairbairn:2017dmf} and \cite{fairbairn2018structure}, have relied on ruling out extremely compact mini structures that collapse from such patches. We caution that it is unclear whether an extrapolation to huge initial overdensity values is physically plausible in realistic models, or whether such patches are sufficiently common to be important. This study restricts to initial isocurvature fluctuations not much greater than unity.

\section{surface density power spectrum}
\label{app:proj}

In this Appendix, we consider a slab with homogeneous and isotropic clumpiness property. We derive the relation between the two-dimensional power spectrum of the surface density field and the three-dimensional power spectrum of the volumetric density field.

We focus in the regime where the slab length $L$ along the direction of projection (i.e. the line of sight) is much larger than the dimensions perpendicular to the direction of projection (i.e. parallel to the plane of the sky). The surface density field can be written as
\ba
\Sigma(\bfr_\perp) = \int^{L/2}_{-L/2} \rmd r_\parallel\,\rho(\bfr) = \int^{L/2}_{-L/2} \rmd r_\parallel\,\int^{+\infty}_{-\infty}\,\frac{\rmd q_\parallel}{(2\pi)}\,\int\,\frac{\rmd^2\bfq_\perp}{(2\pi)^2}\,\tilde{\rho}(\bfq)\,\exp\left[i\left( q_\parallel\,r_\parallel + \bfq_\perp\cdot\bfr_\perp \right)\right].
\ea
Here $r_\parallel$ and $\bfr_\perp$ are the coordinates, in proper units, along and perpendicular to the direction of projection, respectively. The three-dimensional wave vector $\bfq$ can be decomposed into a perpendicular component $\bfq_\perp$ and a parallel component. The Fourier transform of the surface density field is
\ba
\tilde{\Sigma}(\bfq_\perp) = \int^{L/2}_{-L/2} \rmd r_\parallel\,\int^{+\infty}_{-\infty}\,\frac{\rmd q_\parallel}{(2\pi)}\,\tilde{\rho}(\bfq)\,e^{i\, q_\parallel\,r_\parallel}.
\ea
Using $\VEV{\tilde{\rho}(\bfq)\,\tilde{\rho}^*(\bfq')}=(2\pi)^3\,\delta_D(\bfq - \bfq')\,P_\rho(q)$, we can calculate
\ba
\VEV{\tilde{\Sigma}(\bfq_\perp)\,\tilde{\Sigma}^*(\bfq_\perp')} & = & \int^{L/2}_{-L/2} \rmd r_\parallel\,\int^{L/2}_{-L/2} \rmd r'_\parallel\,\int^{+\infty}_{-\infty}\,\frac{\rmd q_\parallel}{(2\pi)}\,\int^{+\infty}_{-\infty}\,\frac{\rmd q_\parallel'}{(2\pi)} \nonumber\\
&& \times (2\pi)^3\,\delta_D(\bfq_\perp - \bfq'_\perp)\,\delta_D(q_\parallel - q'_\parallel)\,P_\rho\left(\sqrt{q^2_\parallel + \bfq^2_\perp} \right)\,e^{i(q_\parallel\,r_\parallel - q'_\parallel\,r'_\parallel)} \nonumber\\
& = & (2\pi)^2\,\delta_D(\bfq_\perp - \bfq'_\perp)\, \int^{L/2}_{-L/2} \rmd r_\parallel\,\int^{L/2}_{-L/2} \rmd r'_\parallel\,\int^{+\infty}_{-\infty}\,\frac{\rmd q_\parallel}{(2\pi)}\,P_\rho\left(\sqrt{q^2_\parallel + \bfq^2_\perp} \right)\,e^{iq_\parallel(r_\parallel - r'_\parallel)} \nonumber\\
& = & (2\pi)^2\,\delta_D(\bfq_\perp - \bfq'_\perp)\,\int^{+\infty}_{-\infty}\,\frac{\rmd q_\parallel}{(2\pi)}\,P_\rho\left(\sqrt{q^2_\parallel + \bfq^2_\perp} \right)\,\left[ \frac{2\,\sin(q_\parallel\,L/2)}{q_\parallel} \right]^2.
\ea
Therefore, the projected power spectrum is
\ba
P_\Sigma(q_\perp) = \int^{+\infty}_{-\infty}\,\frac{\rmd q_\parallel}{(2\pi)}\,P_\rho\left(\sqrt{q^2_\parallel + q^2_\perp} \right)\,\left[ \frac{2\,\sin(q_\parallel\,L/2)}{q_\parallel} \right]^2.
\ea
The regime we are interested in is $q_\perp\,L \gg 1$.

We specialize to the case that $P_\rho(q)\propto q^{\gamma}$ with $\gamma < 1$. In this case, the dominant contribution to the integral comes from modes with $q_\parallel \ll q_\perp$ but $q_\parallel \sim 1/L$. We therefore obtain $P_\Sigma(q_\perp) \sim P_\rho(q_\perp)\,L$. Carrying out the $q_\parallel$-integral explicitly, we can fix the normalization factor of order unity and obtain
\ba
\label{eq:PsigPrho}
P_\Sigma\left( q_\perp \right) & = & P_\rho\left( q_\perp \right)\,L.
\ea
This relation can be verified for a halo model of density inhomogeneity. Under the assumption that halos are uniformly distributed in space, $P_\rho(q)$ and $P_{\Sigma}(q)$ can be explicitly computed in terms of the halo mass function and the halo density profiles, and \refeq{PsigPrho} indeed holds.

\section{Dynamic scales of minihalos}
\label{app:dynamic}

Minihalos that form from the white noise isocurvature density fluctuations have extremely small internal velocity dispersions,
\ba
\label{eq:sigmaapp}
\sigma_v(M) & \sim & \left[ G\,\rho_s(M)\,r^2_s(M) \right]^{1/2} \\
& \simeq & 2 \times 10^{-4}\,{\rm km/s}\,\left( \frac{\nu}{\nu_{\rm med}} \right)^{1/2}\, \left( \frac{\Omega_a}{0.3} \right)^{1/6}\,\left( \frac{1+z_{\rm eq}}{3400} \right)^{1/2}\,\left( \frac{M_0}{10^{-10}\,M_\odot/h} \right)^{1/4}\,\left( \frac{M}{10^{-6}\,M_\odot/h} \right)^{1/{12}}. \nonumber
\ea
Generally, $\sigma_v(M)$ is much smaller than the internal velocities of large CDM halos of any galaxies or galaxy clusters, $\sim 1$--$1000\,{\rm km/s}$. The orbital timescale is
\ba
\label{eq:torbapp}
t_{\rm orb}(M) & \sim & \frac{r_s(M)}{\sigma_v(M)} \\
& \simeq & 30\,h^{-1}\,{\rm Myr}\,\left( \frac{1+z_{\rm eq}}{3400} \right)^{-3/2}\,\left( \frac{\nu}{\nu_{\rm med}} \right)^{-3/2}\,\left( \frac{\Omega_a}{0.3} \right)^{-1/2}\,\left( \frac{M_0}{10^{-10}\,M_\odot/h} \right)^{-3/4}\,\left( \frac{M}{10^{-6}\,M_\odot/h} \right)^{3/4}. \nonumber
\ea
These timescales are much shorter than the age of the present Universe but are usually not so short compared to galactic orbital timescales. Even for minihalos of $M = 10^{-10}\,M_\odot/h$, the dynamic timescale $\sim 30\,$kyr is quite long.

Inside a galaxy cluster, minihalos are flying around at the huge internal velocities of clusters, $\sigma_{\rm cl} \sim 1000\,{\rm km/s}$, taking less than a few years to cross the size of even the largest axion minihalos under our consideration. Therefore, dynamic interactions among minihalos are always much faster than their internal orbital timescales, which is in the regime of {\it high-speed encounters}.

\section{Validity of geometric optics}
\label{app:geoopt}

We show here that in the context of lensing discussed in this paper, geometric optics is valid if observations are done at UV, optical, or near-infrared wavelengths. 
The angular size of a typical star producing the observed micro-caustic crossing events regulates the maximum magnification factor, up to a wavelength at which wave diffraction effects become important. Simple estimates are presented in this Appendix to justify that.

For a single lens plane, the chromatic amplification factor for the amplitude of an electromagnetic wave is given by the following diffraction integral \citep[see, e.g.,][]{Dai:2018enj}
\ba
\label{eq:Flam}
F(\lambda) = \frac{(1+z_L)}{i\,\lambda}\,\frac{D_L\,D_S}{D_{LS}}\,\int\,\rmd^2\bfx\,\exp\left[i\,2\pi\,c\,(1+z_L)\,\tau(\bfx; \bfy)/\lambda \right],
\ea
where $\lambda$ is the wavelength and $\tau(\bfx; \bfy)$ is the ray travel time as a function of the source position $\bfy$ and the image position $\bfx$,
\ba
\tau(\bfx; \bfy) = \frac{D_L\,D_S}{c\,D_{LS}}\,\left[ \frac12\,(\bfx - \bfy)^2 - \phi(\bfx) \right] .
\ea
Considering only the two dominant micro-images during a micro-caustic transit, the Shapiro term $\phi(\bfx)$ can be written as the sum of the contribution from the micro-fold model and a stochastic contribution from surface density fluctuations, $\phi(\bfx) = \phi_{\rm fold}(\bfx) + \Delta\phi(\bfx)$.

Geometric optics is valid if the diffraction integral in \refeq{Flam} can be replaced by complex Gaussian integrals around extrema of $\tau(\bfx;\bfy)$. This requires that $\tau(\bfx;\bfy)$ is locally well approximated by a quadratic function of $\bfx$ in the lowest order Fresnel zones, which in the case of an elongated micro-image with magnification $\mu$, has angular dimensions
\ba
\frac{\Delta x_1}{\sqrt{|\mu|}} \sim \Delta x_2 \sim \left[\frac{\lambda}{2\pi\,(1+z_L)\, D_0}\right]^{1/2} ,
\ea
where we denote $D_0=D_L\,D_S/D_{LS}$.
The validity of the quadratic approximation for $\tau(\bfx; \bfy)$ around extrema is typically limited by the longer dimension $\Delta\,x_1$.

For the micro-fold model, the third derivative of $\tau(\bfx; \bfy)$ is typically comparable to $|\boldsymbol{\tilde d}|$. So the departure from the quadratic approximation for the phase of the integrand in \refeq{Flam} is roughly
\ba
\label{eq:C15}
\left[\frac{\lambda}{2\pi\,(1+z_L)\,D_0}\right]^{1/2}\,\sqrt{|\mu|}\,|\boldsymbol{\tilde d}| & \sim & \frac{10^{-6}\,{\rm rad}}{(1+z_L)^{1/2}}\,\left( \frac{\lambda}{1\,\mu{\rm m}} \right)^{1/2}\,\left( \frac{D_0}{1\,{\rm Gpc}} \right)^{-1/2}\nonumber\\
&& \times \left(\frac{|\mu|}{10^4}\right)^{1/2}\,\left(\frac{|\boldsymbol{\tilde d}|}{10^3\,{\rm arcsec}^{-1}}\right) ~.
\ea
This is negligibly small for reasonable parameters considered in this paper.

If included, small scale surface density fluctuations may be the dominant contribution to the third derivative of $\tau(\bfx; \bfy)$. The characteristic amplitude can be estimated as $\sim \Delta_\kappa/(r_\perp/D_L)$, where the convergence fluctuation has a typical amplitude $\Delta_\kappa$ on a typical transverse proper length scale $r_\perp$. Similar to \refeq{C15}, the correction to the phase of the diffraction integrand is
\ba
\left[\frac{\lambda}{2\pi\,(1+z_L)\,D_0}\right]^{1/2}\,\frac{\sqrt{|\mu|}\,\Delta\kappa\,D_L}{r_\perp} & \sim & \frac{10^{-6}\,{\rm rad}}{(1+z_L)^{1/2}}\,\left( \frac{\lambda}{1\,\mu{\rm m}} \right)^{1/2}\,\left( \frac{D_{\rm eff}}{1\,{\rm Gpc}} \right)^{1/2}\nonumber\\
&& \times \left(\frac{|\mu|}{10^4}\right)^{1/2}\,\left(\frac{\Delta_\kappa}{10^{-4}}\right)\,\left( \frac{r_\perp}{100\,{\rm AU}} \right)^{-1}.
\ea
For a range of parameters relevant to this work, this correction is also unimportant.

\bibliographystyle{aasjournal}
\bibliography{axionmh}



\end{document}